\documentclass[iop]{emulateapj}
\usepackage[dvips]{epsfig}
\usepackage[dvips]{rotating}
\usepackage{amsmath}
\bibpunct{(}{)}{;}{a}{}{,}

\shorttitle{Metal-poor RR\,Lyrae stars in the MCs} 
\shortauthors{Haschke et al.}

\begin{document}

\title{Chemical abundances of metal-poor RR\,Lyrae stars in the Magellanic Clouds\altaffilmark{*}}
\author{Raoul Haschke\altaffilmark{+}, Eva K. Grebel},
\affil{Astronomisches Rechen-Institut, Zentrum f\"ur Astronomie der Universit\"at Heidelberg, M\"onchhofstrasse 12-14, 69120 Heidelberg, Germany}
\and
\author{Anna Frebel} 
\affil{Harvard-Smithsonian Center for Astrophysics, 60 Garden St, MS-20, Cambridge, MA 02138, USA}
\affil{Massachusetts Institute of Technology, Kavli Institute for Astrophysics and Space Research, 77 Massachusetts Avenue, Cambridge, MA 02139, USA}
\and
\author{Sonia Duffau}
\affil{Astronomisches Rechen-Institut, Zentrum f\"ur Astronomie der Universit\"at Heidelberg, M\"onchhofstrasse 12-14, 69120 Heidelberg, Germany}
\and
\author{Camilla J. Hansen \& Andreas Koch}
\affil{Landessternwarte, Zentrum f\"ur Astronomie der Universit\"at Heidelberg, K\"onigstuhl 12, 69117 Heidelberg, Germany}

\slugcomment{submitted: 10 March 2012; accepted: 20 June 2012}

\email{haschke@ari.uni-heidelberg.de}
\altaffiltext{*}{This paper includes data gathered with the 6.5 meter Magellan Telescopes located at Las Campanas Observatory, Chile.}
\altaffiltext{+}{Raoul Haschke is a member of the Heidelberg Graduate School for Fundamental Physics (HGSFP) and of the International Max Planck Research School for Astronomy and Cosmic Physics at the University of Heidelberg.}


\begin{abstract}
We present for the first time a detailed spectroscopic study of chemical element abundances of metal-poor RR\,Lyrae stars in the Large and Small Magellanic Cloud (LMC and SMC). Using the MagE echelle spectrograph at the 6.5\,m Magellan telescopes, we obtain medium resolution ($R \sim 2000 - 6000$) spectra of six RR\,Lyrae stars in the LMC and three RR\,Lyrae stars in the SMC. These stars were chosen because their previously determined photometric metallicities were among the lowest metallicities found for stars belonging to the old populations in the Magellanic Clouds. We find the spectroscopic metallicities of these stars to be as low as $\mathrm{[Fe/H]_{spec}} = -2.7$\,dex, the lowest metallicity yet measured for any star in the Magellanic Clouds. We confirm that for metal-poor stars, the photometric metallicities from the Fourier decomposition of the lightcurves are systematically too high compared to their spectroscopic counterparts. However, for even more metal-poor stars below $\mathrm{[Fe/H]_{phot}} < -2.8$\,dex this trend is reversed and the spectroscopic metallicities are systematically higher than the photometric estimates. 
We are able to determine abundance ratios for ten chemical elements (Fe, Na, Mg, Al, Ca, Sc, Ti, Cr, Sr and Ba), which extend the abundance measurements of chemical elements for RR\,Lyrae stars in the Clouds beyond [Fe/H] for the first time. For the overall [$\alpha$/Fe] ratio, we obtain an overabundance of $0.36$\,dex, which is in very good agreement with results from metal-poor stars in the Milky Way halo as well as from the metal-poor tail in dwarf spheroidal galaxies. Comparing the abundances with those of the stars in the Milky Way halo we find that the abundance ratios of stars of both populations are consistent with another. Therefore we conclude that from a chemical point of view early contributions from Magellanic-type galaxies to the formation of the Galactic halo as claimed in cosmological models are plausible.
\end{abstract}


\defcitealias{Peterson78}{P78}
\defcitealias{Beveridge94}{BS94}
\defcitealias{Clementini95}{C95}
\defcitealias{Casagrande10}{C10}
\defcitealias{Jurcsik95}{J95}
\defcitealias{Zinn84}{ZW84}

\section{Introduction}
\label{Intro}

Hierarchical galaxy formation models predict that larger galaxies are formed by the accretion of smaller systems. Recent simulations \citep[e.g.,][]{Robertson05, Tumlinson10} suggest that the accretion of galaxies with masses comparable to those of the Magellanic Clouds (MCs) provided the major contribution of stellar mass to the present-day Milky Way (MW) halo. These accretion events would have happend early (more than $8-9$\,Gyr ago), contributing primarily stars enriched by Type\,II supernovae \citep[e.g.,][]{Font06}.

The low luminosity dwarf spheroidal (dSph) galaxies contain primarily old populations, which have similar ages as the oldest stars in the MW halo \citep{Grebel04}. 
A number of recent studies succeeded in detecting extremely metal-deficient red giants in Galactic dSphs \citep[e.g.,][]{Frebel10a, Frebel10b, Tafelmeyer10, Cohen09, Cohen10, Norris10a, Norris10b}, concluding that these galaxies may plausibly have contributed to the early build-up of the Galactic halo. For the more metal-deficient red giants in ultra-faint dSph galaxies the similarities to the MW halo are even more pronounced than for the classical dSphs. Therefore it is of particular importance to address the question of the evolution of metallicity in dependence of galaxy mass and type in order to learn more about the similarities and differences between dSphs and dwarf irregular galaxies, such as the MCs, especially in their earliest evolutionary phases. According to the aforementioned cosmological models, the most significant contribution to the build-up of the MW halo should have come from the early progenitors of MC-like galaxies.

In galaxies that experienced star formation throughout cosmic history, the identification of purely old populations is difficult. For instance, the abundant red giants can cover an age range of more than ten Gyr. Thus one can either turn to globular clusters as easily identifiable and reliably age-dateable populations or to uniquely identifiable, old tracer stars among field populations. For field stars the exclusively old Population\,II can only be traced with certain objects, such as RR\,Lyrae stars. Metallicity estimates of these stars therefore allow an insight into the early evolutionary phase of a galaxy. 

The first spectroscopic metallicity estimates of RR\,Lyrae stars in the LMC and SMC were performed by \citet{Butler82}. Using the $\Delta\,S$ method they found mean values of $\mathrm{[Fe/H]} \sim -1.4$\,dex for six LMC RR\,Lyrae stars and $\mathrm{[Fe/H]} \sim -1.8$\,dex for nine SMC RR\,Lyrae stars. Later, low-resolution spectroscopic observations of RR\,Lyrae stars were carried out by \citet{Gratton04} and \citet{Borissova04, Borissova06}, for the LMC. These studies investigated predominantly stars located in the bar and the central regions of the LMC and found very similar mean metallicities\footnote{We assume that the overall metallicity Z of a star is traced by its iron abundance. Throughout this paper, the term metallicity, iron abundance and [Fe/H] are used interchangeably, where $\mathrm{[A/B]} = \mathrm{log}(N_A/N_B) - \mathrm{log}(N_A/N_B)_{\odot}$ for the number N of atoms of elements A and B.} of $\mathrm{[Fe/H]} \sim -1.5$\,dex. The spectroscopic metallicities of RR\,Lyrae stars, determined through comparison of the Ca\,II K line with the hydrogen lines H$_\delta$, H$_\gamma$ and H$_\beta$, were found to be between $\mathrm{[Fe/H]} = -2.33$\,dex and $\mathrm{[Fe/H]} = -0.57$\,dex. These studies could produce overall metallicity estimates only. The abundances of individual elements in RR\,Lyrae stars of the MCs were never investigated.

The Optical Gravitational Lensing Experiment \citep[OGLE\,III,][]{Udalski08a,Udalski08b} monitored the MCs for several years and provided very accurate lightcurves of a large sample of RR\,Lyrae stars \citep{Soszynski09, Soszynski10b}. In \citet{Haschke12_MDF}, the Fourier-decomposition method by \citet{Kovacs95} was used to determine individual photometric metallicity estimates for these old Population\,II stars, and metallicity distribution functions (MDFs) for both MCs were derived. In the LMC, \citet{Haschke12_MDF} obtained Fourier decomposition metallicities for 16776 RR\,Lyrae stars of type\,\textit{ab}. In the SMC metallicities for 1831 of these stars could be measured. On the \citet{Zinn84} scale the peak of the LMC MDF is at $\mathrm{[Fe/H]} = -1.50$\,dex with a full width at half maximum of 0.24\,dex. The SMC is found to be more metal-poor with its MDF peaking at $\mathrm{[Fe/H]} = -1.70$\,dex (FWHM = 0.27\,dex). 

These results are in agreement with \citet{Carrera08a} and \citet{Carrera11}, who used Ca\,II triplet spectroscopy investigating red giants in the LMC to determine the age-metallicity relation, as well as with metallicities from Ca\,II triplet spectroscopy for SMC red giants determined by \citet{Carrera08b}. These studies estimate the mean metalliticy of the \textquotedblleft old\textquotedblright \ LMC population as traced by field red giants to be $\mathrm{[Fe/H]} \sim -1.2$\,dex and that of the SMC to be $\mathrm{[Fe/H]} \sim -1.4$\,dex. Since the majority of the red giants in both clouds were formed at ages much younger than 10\,Gyr, these metallicities are actually representative of the intermediate-age population. The age determination for individual field red giants, however, is rather difficult, and is exacerbated by uncertainties in the distance and the reddening of these stars. In contrast, star clusters usually have a well defined age with a common metallicity for all stars. \citet{Kayser07} and \citet{Glatt08b} found a considerable metallicity spread for SMC star clusters of intermediate age based on Ca\,II triplet spectroscopy, in agreement with the spectroscopic results by \citet{DaCosta98}, and concluded that no smooth, monotonic age-metallicity relation is present. 

Hence, the only source of direct element abundance measurements for purely old stars of the MCs were studies of globular clusters (GCs) so far. In the LMC seven old GC were spectroscopically investigated by \citet{Johnson06} and \citet{Mucciarelli09,Mucciarelli10} using red giants. They found that the different LMC GCs studied have mean iron abundances between $-2.20\,\mathrm{dex} < \mathrm{[Fe/H]} < -1.23$\,dex. These findings support that a considerable metallicity spread is present in the old cluster population. While the GCs of \citet{Mucciarelli10} are all $\alpha$-enhanced by $\sim 0.4$\,dex, the GCs investigated by \citet{Johnson06} are only slightly $\alpha$-enhanced. However, the abundances of the $\alpha-$ and other elements agree well with the abundances determined for stars within the MW halo. The SMC only contains one GC and this object is $2-3$\,Gyr younger than the oldest GCs in the LMC and in the MW \citep{Glatt08a}.

But even for Galactic field RR\,Lyrae stars hardly any data are available concerning the $[\alpha/\mathrm{Fe}]$ ratio. \citet{Clementini95} reported an overabundance of $[\alpha/\mathrm{Fe}] = 0.4$\,dex and $\mathrm{[Mn/Fe]} = 0.6$\,dex for 10 Galactic field RR\,Lyrae stars, which is in agreement with \citet{Fernley96}, who found an overabundance of $\mathrm{[Ca/Fe]} = 0.4$\,dex for nine field stars. In the Milky Way GC M3 \citet{Sandstrom01} measured a similar overabundance using 29 RR\,Lyrae stars. Two very metal-poor RR\,Lyrae stars were investigated with high resolution spectroscopy by \citet{Hansen11}. They obtained elemental abundances for 16 different element and found the $[\alpha/\mathrm{Fe}]$ ratio to be very similar to the other studies. \citet{For11} investigated the properties of eleven nearby field RR\,Lyrae stars and obtained very precise abundances for several elements. For the $\alpha$-elements, they found a mean overabundance of $[\alpha/\mathrm{Fe}] = 0.5$\,dex.

This lack of information about $\alpha$-elements for LMC RR\,Lyrae stars and the lack of spectroscopic investigations of SMC RR\,Lyrae stars is caused in part by the great difficulties of taking spectra of comparatively faint objects at these distances. The variability of these pulsating stars changes their stellar parameters significantly during a given pulsation period. Therefore the integration times have to be kept short. At distances beyond $\sim 50$\,kpc, only the $>$\,6-m class telescopes and state-of-the-art instruments are able to obtain spectra of sufficient quality to permit us to determine the chemical composition of these stars. Therefore, for the time being, photometric estimates are the only way to determine metallicities for large samples of RR\,Lyrae stars. Photometric metallicity calibrations at the low metallicity end, however, are still somewhat uncertain \citep[e.g.,][]{Arellano11, Dekany09}, owing to the lack of suitable calibration stars. To arrive at more secure metallicity estimates of metal-poor candidates, spectroscopy is essential. 

Of particular interest are the very metal-poor and thus presumably oldest populations of the MCs to learn more about the earliest stages of their evolution and to compare the early enrichment history of the MCs with the properties of the Galactic halo and other galaxies orbiting the MW. In this study, we therefore chose to conduct spectroscopic follow-up observations of six LMC RR\,Lyrae stars and three SMC RR\,Lyrae stars that were previously identified as very metal-poor candidates in our photometric analysis. Their photometric metallicities, deduced from the method introduced by \citet{Kovacs95} are all below $\mathrm{[Fe/H]} < -2.3$\,dex on the scale of \citet{Zinn84}. For more details on the metallicities of RR\,Lyrae stars within the MCs see \citet{Haschke12_MDF}. 

In Section\,\ref{Observations}, the target selection and the observational procedure to obtain spectra for these stars is described. Furthermore, the reduction procedure of the data is explained. We measure the radial velocities of the RR\,Lyrae stars in Section\,\ref{RadVel} and correct the spectra accordingly to obtain equivalent width measurements in Section\,\ref{EquWidth}. The stellar parameters are deduced and discussed in Section\,\ref{Parameter}. Equivalent width measurements as well as synthesized spectra are used in Section\,\ref{Abundances} to calculate abundances for eleven different elements. In Section\,\ref{Conclusions} the results are discussed and summarized.


\section{Observations and Reductions}
\label{Observations}

\subsection{Target selection}

The OGLE\,III survey\footnote{\url{http://ogle.astrouw.edu.pl/}} \citep{Udalski08b} monitored the MCs from 2001 to 2009. \citet{Soszynski09, Soszynski10b} extracted very precise light curves for several thousand RR\,Lyrae stars in both Clouds. These authors then Fourier-decomposed these light curves, following the method by \citet{Kovacs95}, to determine their Fourier parameters. \citet{Smolec05} introduced a relation using the Fourier parameter $\phi_{31}$ and the period of the RR\,Lyrae stars to determine a metallicity on the metallicity scale of \citet{Jurcsik95}. In \citet{Haschke12_MDF}, this relation is used to evaluate photometric metallicities for 16676 LMC RR\,Lyrae stars and 1831 SMC RR\,Lyrae stars. \citet{Papadakis00} found a transformation relation between the metallicity scales by \citet{Jurcsik95} and \citet{Zinn84}, showing that the scale by \citet{Jurcsik95} is about 0.3\,dex higher than the metallicity scale by \citet{Zinn84}. 

For the LMC, our candidate list contains 35 RR\,Lyrae stars of type\,{\em ab} with photometric metallicities below $\mathrm{[Fe/H]} = -2.0$\,dex on the \citet{Jurcsik95} scale, which scales to $\mathrm{[Fe/H]} = -2.3$\,dex on the scale of \citet{Zinn84}. When we compiled the candidate list for the SMC spectroscopic observations the data on the RR\,Lyrae stars from OGLE\,III were not available. We therefore had to rely on OGLE\,II data \citep{Udalski97, Soszynski03}, which result in 14 RR\,Lyrae type\,\textit{ab} stars with metallicities $\mathrm{[Fe/H]_{ZW84}} < -2.3$\,dex.

\subsection{Observations}

The observations were conducted using the Magellan Echellette (MagE) spectrograph \citep{Marshall08} on the 6.5\,m Magellan Clay telescope at Las Campanas, Chile, on 2010 January 17 and 18 and August 4 and 5. The spectrograph has a total wavelength coverage from 3000\AA \ to 10000\AA.

RR\,Lyrae stars have periods of $12-20$ hours. Due to their pulsation their stellar parameters, such as temperature and gravity, vary significantly in a relatively short amount of time \citep[e.g.,][]{Layden94}. To obtain a useful spectrum the integration time should not exceed 1.5\,hours to avoid broadening of the lines. Due to the parameter variability the observations should be conducted within the phase interval of $\Phi = 0.2 - 0.8$, where $\mathrm{\Phi} = 0$ and $\mathrm{\Phi} = 1$ are two consecutive maxima of the light curve. This interval of $\Phi = 0.2 - 0.8$ corresponds to the time of minimum brightness and activity of the star. Hence, we always calculated the actual $\Phi$ before starting the observation, in order to ensure to observe at the right phase. 

\begin{figure}
\includegraphics[width=0.47\textwidth]{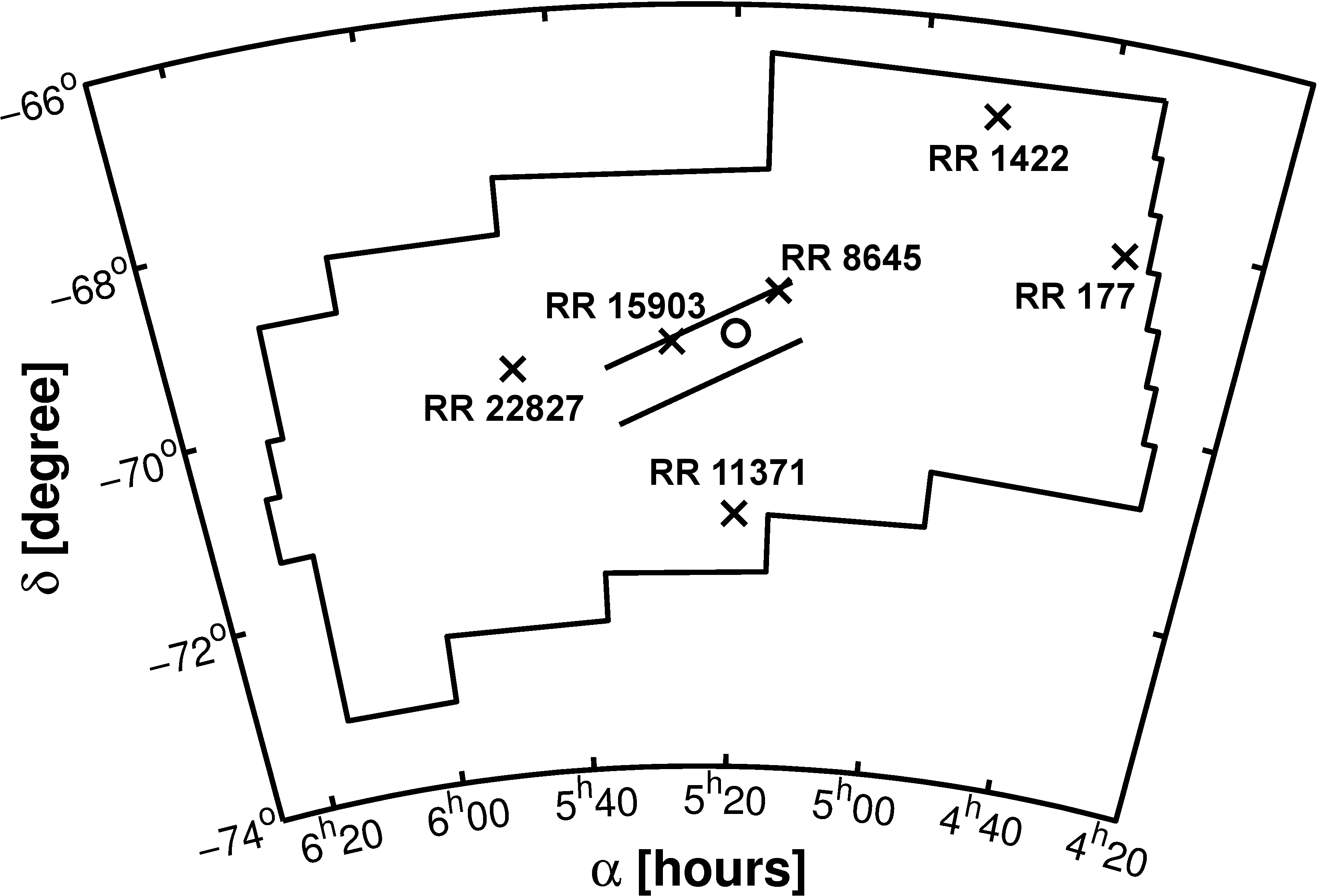} 
\vspace{\floatsep}
\includegraphics[width=0.47\textwidth]{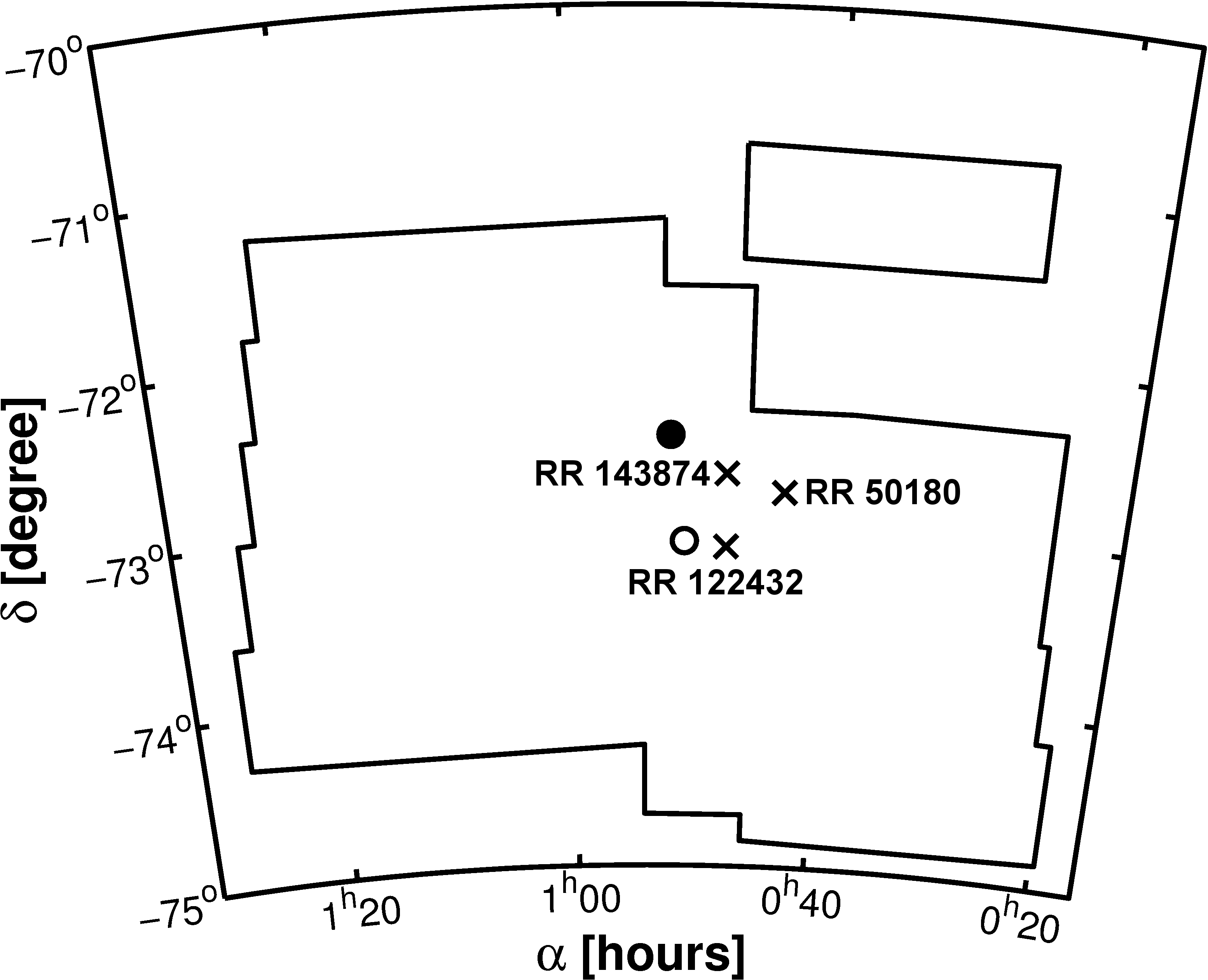} 
\caption{Location of the observed RR\,Lyrae stars within the OGLE\,III fields. The crosses illustrate the position of the RR\,Lyrae stars. The open circle in the upper panel shows the optical center of the LMC \citep{Vaucouleurs72}, while the two diagonal lines indicate the position of the LMC bar. The open circle in the lower panel shows the center of the SMC as found by \citet{Gonidakis09} from isophotes of giant stars, while the filled circle indicates the center from proper motion measurements as found by \citet{Piatek08}.}
\label{location_RRL}
\end{figure}

The list of observed stars is shown in Table\,\ref{list_RRL} and Figure\,\ref{location_RRL}. Due to changing seeing conditions, we used a $0.7\arcsec$ slit for the spectra of the first three LMC targets and all of the SMC targets and a $2.0\arcsec$ slit for the other three LMC RR\,Lyrae stars. With the $0.7\arcsec$ slit, a resolving power of $R \equiv \lambda/\Delta\lambda \sim 6000$ can be achieved, while the $2.0\arcsec$ slit has a resolving power of $R \sim 3000$. The integration time for each object was between $45 - 90$\,min, depending on the slit width and the brightness of the target. The analysis for the LMC target RR8645 was conducted with special care, because a second star was present in the slit during the observation.

To calibrate the wavelength solution several ThAr-lamp exposures were taken throughout the night. Also at the beginning and the end of each night flat fields were taken.

\begin{table*}
\caption{Properties of RR\,Lyrae stars observed with Magellan/MagE. The star ID of the LMC objects is taken from the identifier in the OGLE\,III dataset. For the SMC objects the OGLE\,II identifier is used.}             
\label{list_RRL}      
\centering                          
\begin{tabular}{lcccccccccc}        
\hline\hline                 
star & V & period & RA\,(J2000) & Dec\,(J2000) & UT date & No. integrations & $t_{\mathrm{exp}}$ & slit & [Fe/H]$_{\mathrm{phot}}$ & Blazhko effect \\
     & [mag] & [days] & [hh:mm:ss] & [$^\circ$:\arcmin:\arcsec] & &  & [s] & [\arcsec] & [dex] & present \\
\hline                        
RR15903 & 18.78 & 0.7453036 & 05:27:18.77 & $-69$:32:22.5 & 01/18/2010 & 4 & 4800 & 0.7 & $-2.20$ & no \\ 
RR8645 & 19.23 & 0.5293966 & 05:14:42.54 & $-68$:59:26.9 & 01/18/2010 & 3 & 5400 & 0.7 & $-2.63$ & no \\
RR1422 & 19.11 & 0.6148814 & 04:51:54.23 & $-66$:58:46.1 & 01/18/2010 & 2 & 5100 & 0.7 & $-3.38$ & yes\\ 
RR177 & 18.84 & 0.8246837 & 04:35:33.16 & $-68$:12:25.8 & 01/19/2010 & 3 & 2700 & 2.0 & $-2.10$ & no \\
RR11371 & 19.27 & 0.5590900 & 05:19:27.68 & $-71$:20:45.0 & 01/19/2010 & 2 & 3600 & 2.0 & $-2.80$ & yes\\
RR22827 & 19.16 & 0.5720020 & 05:46:40.62 & $-69$:44:26.1 & 01/19/2010 & 1 & 3000 & 2.0 & $-2.80$ & yes\\

RR122432 & 19.63 & 0.590461 & 00:47:39.98 & $-73$:08:43.0 & 08/05/2010 & 3 & 4100 & 0.7 & $-3.41$ & yes\\
RR50180 & 20.17 & 0.585893 & 00:43:07.78 & $-72$:49:22.1 & 08/06/2010 & 2 & 3000 & 0.7 & $-2.95$ & yes\\
RR143874 & 20.21 & 0.552044 & 00:47:42.57 & $-72$:43:15.5 & 08/06/2010 & 3 & 5600 & 0.7 & $-2.03$ & yes \\
\hline                                   
\end{tabular}
\end{table*}

Three metal-poor standard stars of known metallicity were observed for comparison: the RR\,Lyrae star X\,Arietis (X\,Ari) and the non-variable stars HD74000 and G64-12. Using the $0.7\arcsec$ slit we took several exposures of a few seconds for every standard star. For the analysis, the individual spectra of a single star were co-added. For the MC RR\,Lyrae stars a final S/N level of $10 - 25$ per pixel was achieved for the combined observations at a wavelength of 7500\AA. For the bright standard stars the S/N level is much better, with values between $150 - 200$ per pixel. The stellar parameters of these stars are shown in Table\,\ref{list_standards}.

\begin{table*}
\caption{Properties of standard stars observed with Magellan/MagE. The values listed are taken from the quoted literature.} 
\label{list_standards}
\centering                          
\begin{tabular}{lccccccccc}        
\hline \hline
Star & V & RA\,(J2000) & Dec\,(J2000) & [Fe/H] & T$_{\mathrm{eff}}$ & log\,g & Reference \\
 & [mag] & [hh:mm:ss] & [$^\circ$:\arcmin:\arcsec] & [dex] & [K] &  &  & \\ 
\hline
HD74000 & 9.67 & 08:40:50.80 & $-16$:20:42.52 & $-2.02$ & 6166 & 4.19 & \citet{Cenarro07} \\
X\,Ari & 9.63 & 03:08:30.88 & $+10$:26:45.22 & $-2.50$ & 6109 & 2.60 & \citet{Cenarro07} \\
GD64-12 & 11.49 & 13:40:02.49 & $-00$:02:18.75 & $-3.28$ & 6400 & 4.1 & \citet{Primas00} \\
\hline
\end{tabular}
\end{table*} 

\subsection{Data Reduction}

The low S/N level of the data requires a careful reduction procedure. We chose to try two different approaches, reducing the data by hand using the IRAF packages and an IDL based pipeline. The results are compared to obtain the best results possible. 

\begin{figure}
\includegraphics[width=0.47\textwidth]{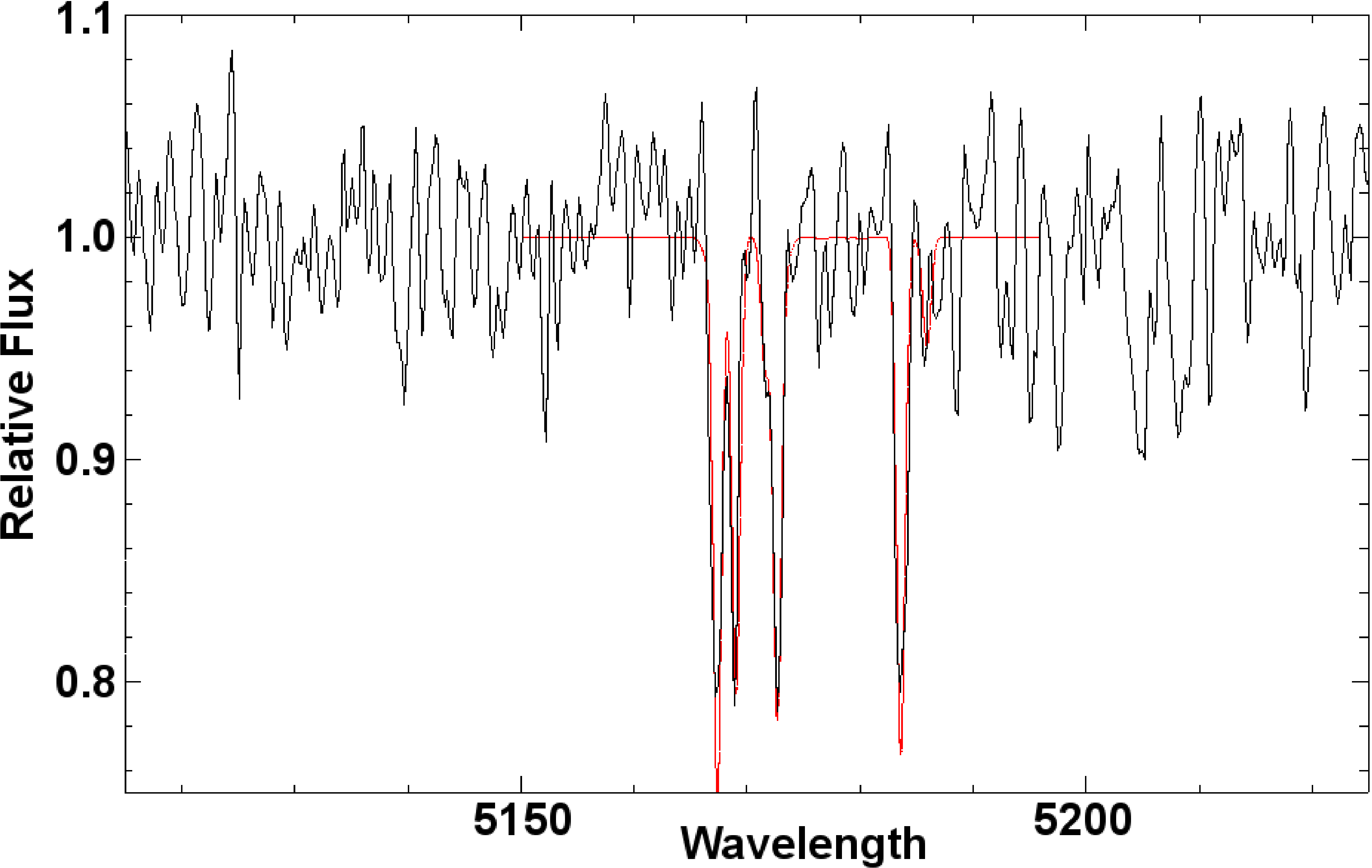} 
\caption{Spectrum of RR1422 at the region of the Mg lines around 5170\AA. The red line indicates the best fitting synthesized model.}
\label{Spectra_RR1422}
\end{figure}

Finally we decided to use the IDL pipeline, written by G. Becker\footnote{\url{ftp://ftp.ociw.edu/pub/gdb/mage\_reduce/mage\_reduce.tar.gz}}, to reduce the data. The pipeline provides the user with a wavelength-calibrated spectrum. With standard IRAF routines we normalize the spectrum and merge the different orders of each spectrum to obtain a final one-dimensional spectrum. An exemplary spectrum is shown in Figure\,\ref{Spectra_RR1422}, spectra of the other target and standard stars are shown in Figure\,\ref{spectra_Mg_standards_SMC} and Figure\,\ref{spectra_Mg_LMC} in the appendix.

\subsection{Uncertainties}
\label{Uncertainties}

For the discussion of our uncertainties we will always quote the standard deviations 
\begin{equation}
\sigma = \sqrt{\frac{1}{n} \Bigl{(} \sum_{n = 1}^{i}(x_i-\mu)^2) \Bigr{)} \frac{n}{n-1}}
\end{equation}
where $n$ is the number of measurements, $x_i$ are the individual datapoints and $\mu$ is the mean of the sample. The last term $n/(n-1)$ corrects for the effects of small number statistics. We do not use the standard error $\sigma/\sqrt{n}$, unless explicitly stated.


\section{Radial Velocities}
\label{RadVel}

The pulsation of an RR\,Lyrae star causes periodic changes in the observed spectra and hence of the measured radial velocity. For RR\,Lyrae type\,{\em ab} stars, the radial velocity changes by up to $60-70$\,km\,s$^{-1}$ during a given period, with the lowest velocity around minimum light \citep{Smith04}. Moreover, the different layers of the stellar atmosphere move with a velocity that gradually increases towards the outer shells. Since metallic and Balmer lines originate at different optical depths in the atmospheres, the pulsation can introduce systematic velocity offsets in the lines used for the velocity estimate depending at which phase the spectra were taken. 

We determine the radial velocities from the Doppler shift of the near-infrared Ca\,II-triplet lines between 8492\AA \ and 8662\AA \ and of the Balmer lines, H$\gamma$, H$\beta$ and H$\alpha$, between 4333\AA \ and 6568\AA. Taking all individual measurements, we compute the mean and median value of the radial velocity as well as the standard deviation. 

Using the approach described in \citet{Vivas05}, we estimate the pulsational velocity of each target star to obtain the systemic velocity. The template pulsational velocity curve of X\,Ari, observed by \citet{Oke66} and parameterized by \citet{Layden94}, is adopted for all target RR\,Lyrae stars. The velocity curve of X\,Ari is fitted at the observed phase of our targets to derive the corresponding pulsational velocity. However, for each star only measurements at one certain phase of the period are available and a fit to the radial velocity curve is rather uncertain. To calculate the uncertainties of the systemic velocities we adopt Equation\,1 of \citet{Vivas05}

\begin{equation}
\sigma^{2}_{\gamma} = \sigma^{2}_{r} + (119.5 \times 0.15)^2 + (23.9\Delta\Phi)^2,
\label{error_systemic_velocity}
\end{equation}

where $\sigma_{\gamma}$ is the uncertainty of the systemic velocity and $\sigma_{r}$ the standard deviation of the radial velocity measurements. The second term reflects the duration of the observation of the spectra. This took usually about 15\% of the target's phase. This duration of the observation is convolved with the slope of the velocity curve of 119.5\,km\,s$^{-1}$. The third term reflects the 1$\sigma$ star-to-star variation of the velocity amplitude of 23.9\,km\,s$^{-1}$, as found by \citet{Vivas05}, combined with the difference of the observed phase of our target star to the phase with zero systemic velocity, which is found to occur at $\Phi = 0.5$ for X\,Ari.

In Table\,\ref{RV} we present the median heliocentric systemic radial velocities, as well as systemic radial velocities in the local standard of rest (LSR), using the IRAF task \textit{rvcorrect}. The mean velocities are very similar to the median velocities in Table\,\ref{RV}. Additionally the pulsational velocity of the star, the mean phase of the observation and the standard deviation of the radial velocity measurements as well as the uncertainty of the systemic velocity are shown.

\begin{table}
\caption{Various velocity estimates and uncertainties of all our target stars and X\,Ari} 
\label{RV}
\centering                          
\begin{tabular}{lrrrrrr}        
\hline \hline
Star & V$_{\mathrm{helio}}$ & V$_{\mathrm{LSR}}$ & mean & $\sigma_{r}$ & $\sigma_{\gamma}$ \\
 & [km\,s$^{-1}$] & [km\,s$^{-1}$] & phase & [km\,s$^{-1}$] & [km\,s$^{-1}$]   \\ 
\hline
RR15903 & 320 & 304 & 0.64 & 5 & 19 \\
RR8645 & 292 & 277 & 0.38 & 23 & 29 \\
RR1422 & 325 & 309 & 0.57 & 11 & 21 \\
RR177 & 287 & 271 & 0.18 & 20 & 27 \\
RR11371 & 323 & 308 & 0.41 & 30 & 35 \\
RR22827 & 295 & 279 & 0.81 & 33 & 38 \\
RR122432 & 174 & 163 & 0.45 & 33 & 38 \\
RR50180 & 133 & 122 & 0.66 & 18 & 25 \\
RR143874 & 109 & 98 & 0.41 & 8 & 20 \\
X\,Ari & $-40$ & $-51$ & 0.51 & 15 & 23 \\
\hline
\end{tabular}
\end{table} 

The values for the heliocentric radial velocity of X\,Ari found by us are in very good agreement with the literature, for instance, \citet{Valdes04} found a value of $-40$\,km\,s$^{-1}$. 

The mean systemic heliocentric velocity of our small sample of LMC RR\,Lyrae stars is found to be $307 \pm 19$\,km\,s$^{-1}$. This value is more than $1\,\sigma$ greater than the radial velocities of \citet{Gratton04}, who found $261 \pm 40$\,km\,s$^{-1}$, of \citet{Borissova04}, who found $264 \pm 19$\,km\,s$^{-1}$, or of \citet{Borissova06} who found $255 \pm 65$\,km\,s$^{-1}$. However, these measurements were made in the most central parts of the LMC and it seems reasonable that the disk region may have different velocities compared to the central regions in and close to the bar. 


In the sample of \citet{Borissova06}, some high-velocity stars with radial velocities up to 399\,km\,s$^{-1}$ are present. We compare their metallicities of these high-velocity stars with their metallicities for the other stars. Both groups of stars have an indistinguishable mean metallicity and therefore we do not find any correlation between velocity and metallicity.

Even though, the radial velocity of the LMC seems very similar for different stellar tracers, the velocity dispersion increases with the age of the stellar tracer. With a sample of 738 red supergiants distributed within several degrees from the center of the LMC, \citet{Olsen11} found a systemic velocity of $263 \pm 2$ \,km\,s$^{-1}$. A mean radial velocity of 257\,km\,s$^{-1}$ was estimated by \citet{Cole05} using red giants. The velocity dispersion of the LMC has been measured to 9\,km\,s$^{-1}$ for young red supergiants \citep{Olsen07}, while old RR\,Lyrae stars have a velocity dispersion of about 50\,km\,s$^{-1}$ \citep{Minniti03, Borissova06}. For intermediate-age tracers a gradual increase is found \citep[compare, e.g.,][]{Marel09}.

One possibility for the high velocities of our sample is an effect of small number statistics. Another possibility is that our metal-poor RR\,Lyrae stars trace an older, earlier population than the more metal-rich RR\,Lyrae stars studied before, possibly tracing a sparse, metal-poor halo.

For our three RR\,Lyrae stars in the SMC we find a mean systemic velocity of $138 \pm 40$\,km\,s$^{-1}$. In the literature no velocity estimates were carried out for RR\,Lyrae stars so far. Only samples of brighter SMC stars were used to determine the distribution of the radial velocity. With more than 2000 red giant branch stars, \citet{Harris06} found a Gaussian distribution of the velocity with a maximum at $145.6 \pm 0.6$\,km\,s$^{-1}$ and a velocity dispersion of $27.6 \pm 0.5$\,km\,s$^{-1}$. Our results are in good agreement with these values. \citet{DePropris10} evaluated red giants in ten different fields with radial distances of 0.9\,kpc to 5.1\,kpc from the SMC center and found that the velocity distribution functions between the different fields are significantly different. Some fields show single peak distributions, while others seem to be bimodal. The mean radial velocity changes also between the fields and they find values between $147 \pm 26$\,km\,s$^{-1}$ and more than 200\,km\,s$^{-1}$ for the second peak in fields with a bimodal distribution. Overall, our velocity estimates are in good agreement with the velocities of \citet{DePropris10}. However, our stellar sample for the SMC needs to be enlarged first before more detailed conclusions can be drawn.


\section{Equivalent Width Measurements}
\label{EquWidth}

Our line list for the investigated RR\,Lyrae stars of the MCs is primarily based on the line list of \citet{Clementini95}. They investigated, among others, X\,Ari, which we use as a standard RR\,Lyrae star in our analysis. The line list of \citet{Clementini95} is complemented by the lists of \citet{Peterson78} and \citet{Beveridge94}, who obtained abundances for HD\,74000, the other standard star in our investigation. Before we employ the merged list, we test for blends by comparing our list with the line list extracted from the Vienna Atomic Line Database (VALD), set up by \citet{Piskunov95}. All blended lines are removed from the sample.

With the IRAF task \textit{splot}, a Gaussian is fitted to the line profile of every unblended atomic line in our line list. We remove all lines that are obviously contaminated by noise or cosmic rays. For our science targets, very strong lines (EW $> 200$\,m\AA) as well as weak lines (EW $< 20$\,m\AA) were excluded from further analysis. The strong lines might be saturated, while for the weak lines the uncertainties of the EW measurements are too large due to the low S/N. The remaining values for the EWs of our target stars, together with the literature values for the standard stars, the excitation potential ($\chi$) and the value for the oscillator strength (log\,gf), are given in Table\,\ref{EW_table}. 

\begin{table}
\caption{Equivalent width measurements for all target and standard stars.} 
\label{EW_table}
\centering                          
\begin{tabular}{lccrc}        
\hline \hline
Element & $\lambda$ & $\chi$ & log\,gf & HD\,74000$_{\mathrm{H11}}$  \\
 & [\AA] & [eV] &  & [m\AA] \\  
\hline
Na\,I & 5889.97 & 0.00 & $0.108$ & 175 \\
Na\,I & 5895.94 & 0.00 & $-0.194$ & 130 \\
Mg\,I & 4057.52 & 4.34 & $-0.890$ & \nodata \\
Mg\,I & 4167.28 & 4.34 & $-1.010$ & 31 \\
\hline
\end{tabular}
\tablecomments{Only the first few lines and columns are shown. The full table is available electronically.}
\end{table} 

Comparing our EW measurements of the standard star HD\,74000 we find very similar results to the literature values by \citet{Peterson78} with a mean difference of $\Delta\mathrm{EW} = 5.9$\,m\AA \ and a standard deviation of $16.4$\,m\AA. Even better agreement is achieved by comparing the measurements to \citet{Beveridge94}. A mean difference of $\Delta\mathrm{EW} = 3.5$\,m\AA \ with a standard deviation of $7.6$\,m\AA \ is obtained. The literature values of the EWs of X\,Ari obtained by \citet{Clementini95} are generally higher by $\Delta\mathrm{EW} = 38.6$\,m\AA \ than the measurements by us. We explain this rather large difference with a different observed phase of the variable RR\,Lyrae star X\,Ari. While \citet{Clementini95} observed X\,Ari between $0.71 < \Phi < 0.82$, our observations were conducted between $0.47 < \Phi < 0.55$. The physical conditions, such as a temperature and surface gravity, are therefore altered.


\section{Stellar Parameters}
\label{Parameter}

For the analysis of the chemical abundances of our nine RR\,Lyrae stars in the MCs as well as for the three standard stars, we use two independent approaches. In order to determine the stellar abundances either from EWs or spectral synthesis we need to know the following parameters: effective temperature ($T_{eff}$), surface gravity (log\,g), metallicity ([Fe/H]), and microturbulence ($\nu_{\mathrm{t}}$). Interpolation of Kurucz's atmosphere model grid without convective overshooting \citep{Castelli97} is used to construct model atmospheres for the different stars. The local thermodynamic equilibrium (LTE) spectral line synthesis code MOOG\footnote{http://www.as.utexas.edu/$\sim$chris/moog.html} \citep{Sneden73} is used, in its version from 2010, to derive the elemental abundances from the EW measurements and from the synthesis of the spectra. We assume the solar and meteoritic abundances from \citet{Anders89}, except for the iron content of log\,$\varepsilon$(Fe) = 7.52, taken from the solar abundance catalogue of \citet{Grevesse98}.

\subsection{Effective Temperature}

\citet{Clementini95} derived a relation to calculate the effective temperature $T_{eff}$ of an RR\,Lyrae star from the differences of its brightness in different filters. Their Equation\,4 summarizes the relations for four different colors. In OGLE\,III, the $V$ and $I$ filter are used. Therefore, we use 

\begin{eqnarray}
T_\mathrm{eff/C95} = 753.906677(V-I)^2 \label{eff_T_C95}\\
 - 4836.16016(V-I) + 8801.4248.\nonumber
\end{eqnarray}

Investigating dwarfs and subgiants \citet{Casagrande10} found a relation for the reciprocal effective temperature, $\theta_\mathrm{eff} \equiv 5040/T_\mathrm{eff}$, based on the color of the star and its metallicity [Fe/H]. This equation has not been tested for variable stars yet. We want to test whether the relation is applicable and adopt their Equation\,3

\begin{eqnarray}
&\theta_\mathrm{eff/C10} = 0.4033 + 0.8171(V-I) - 0.1987(V-I)^2 \label{eff_T_C10}\\
 &- 0.0409(V-I)\mathrm{[Fe/H]} + 0.0319\mathrm{[Fe/H]} + 0.0012\mathrm{[Fe/H]}^2. \nonumber
\end{eqnarray}

To calculate the effective temperature of each star at the phase of observation, we need the color information for this specific time. The lightcurve of the RR\,Lyrae star is recovered using the individual observations taken over several years by the OGLE\,III collaboration in the $V$ and preferentially the $I$\,band. As a first approach, we fit the data points of the individual observations with a tenth order polynomial for the $I$\,band and a fifth order polynomial for the $V$\,band. The data in the $I$\,band are much better sampled because of the many repeat observations in this filter, while the fit of the $V$\,band is more susceptible to the influence of outliers. Especially the phase of rapid atmospheric change at $\Phi = 0.8 - 0.2$ is not very well sampled and therefore the fitted lightcurve is not reliable in this range. 

As a second approach to determine the lightcurve in the $V$\,band, we use the six template light curves \citet{Layden98} introduced, based on his study of 103 Galactic RR\,Lyrae type\,\textit{ab} stars. By using a $\chi^2$ test, we find the template light curve fitting our target stars best.

\begin{figure}
\includegraphics[width=0.47\textwidth]{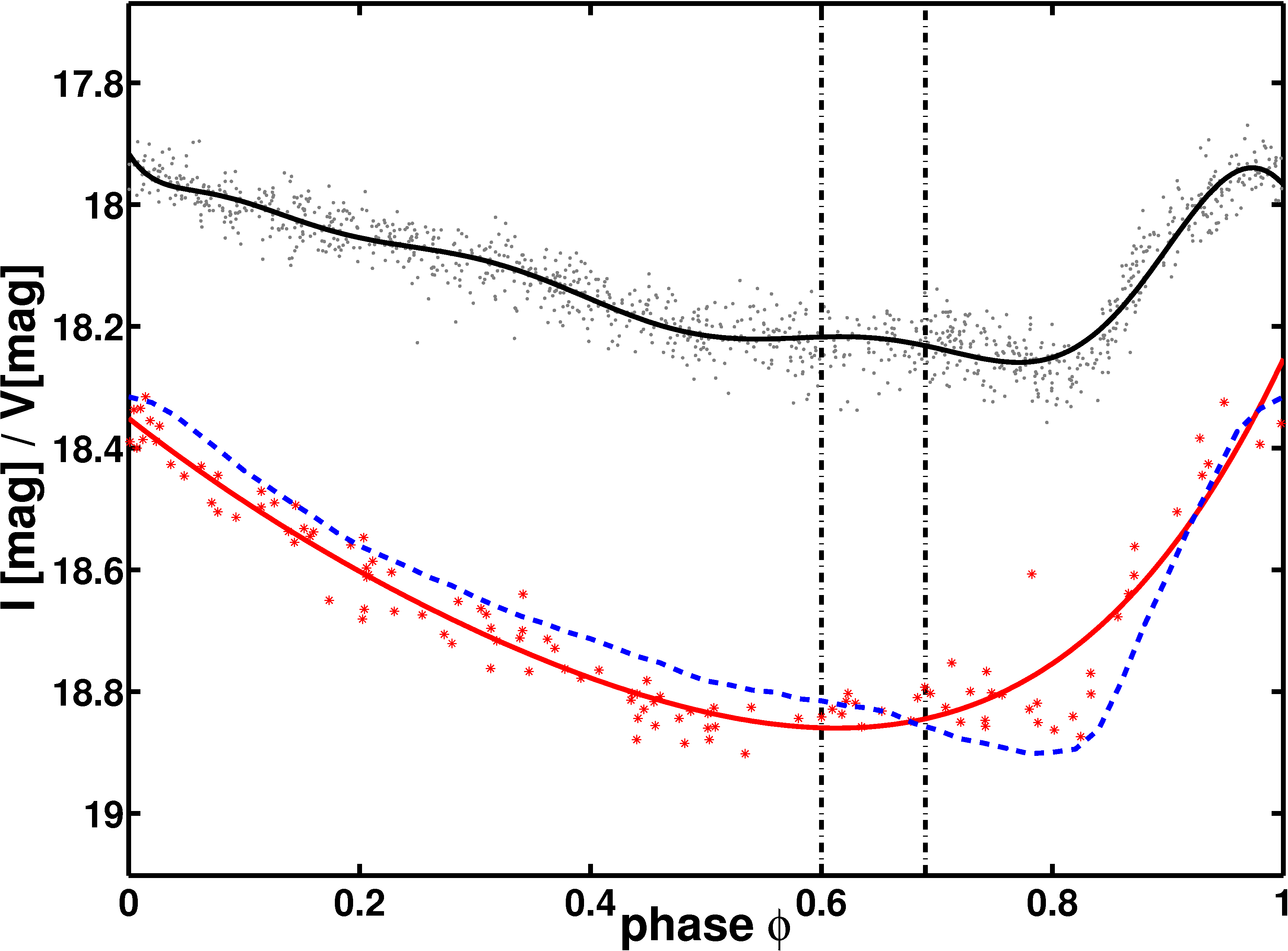} 
\caption{An exemplary light curve for our target stars, here RR\,15903. The grey dots are the observed $I$\,band magnitudes, while the black line is the corresponding fit. $V$\,band magnitudes are shown with red asterisks. The red line is the polynomial fit. The best template of \citet{Layden98} is shown with the blue dotted line. The vertical black dash-dotted lines indicate the actually used observation window for RR15903.}
\label{RR_lightcurve}
\end{figure}

Subtracting the $I$\,band light curve (black line in Figure\,\ref{RR_lightcurve}) from the $V$\,band light curve from the polynomial fit (red line in Figure\,\ref{RR_lightcurve}), as well as from the template light curve of \citet{Layden98} (blue dotted line in Figure\,\ref{RR_lightcurve}), we obtain two sets of colors for the different phases of the RR\,Lyrae stars. 

\begin{figure}
\includegraphics[width=0.47\textwidth]{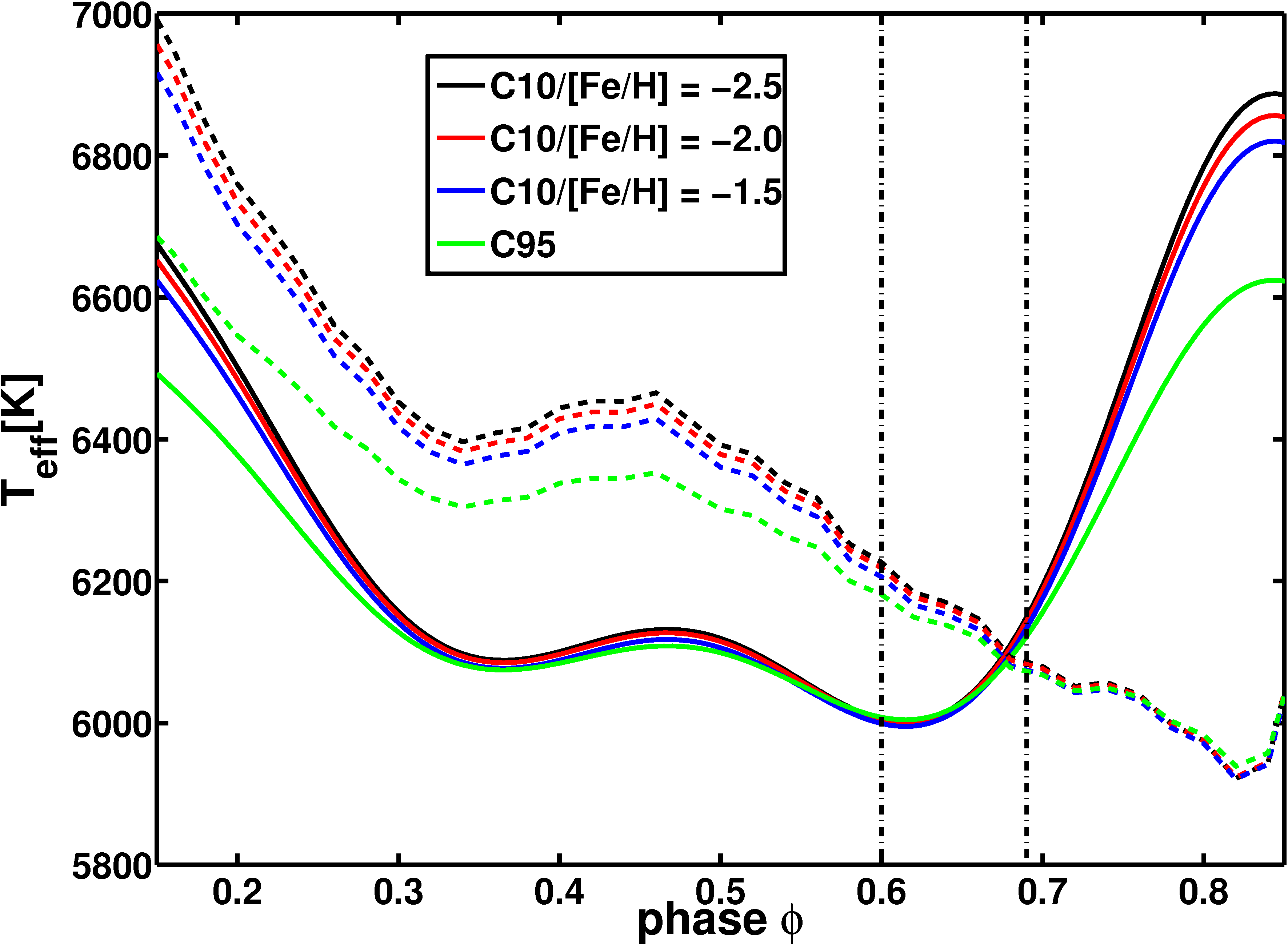} 
\vspace{\floatsep}
\includegraphics[width=0.47\textwidth]{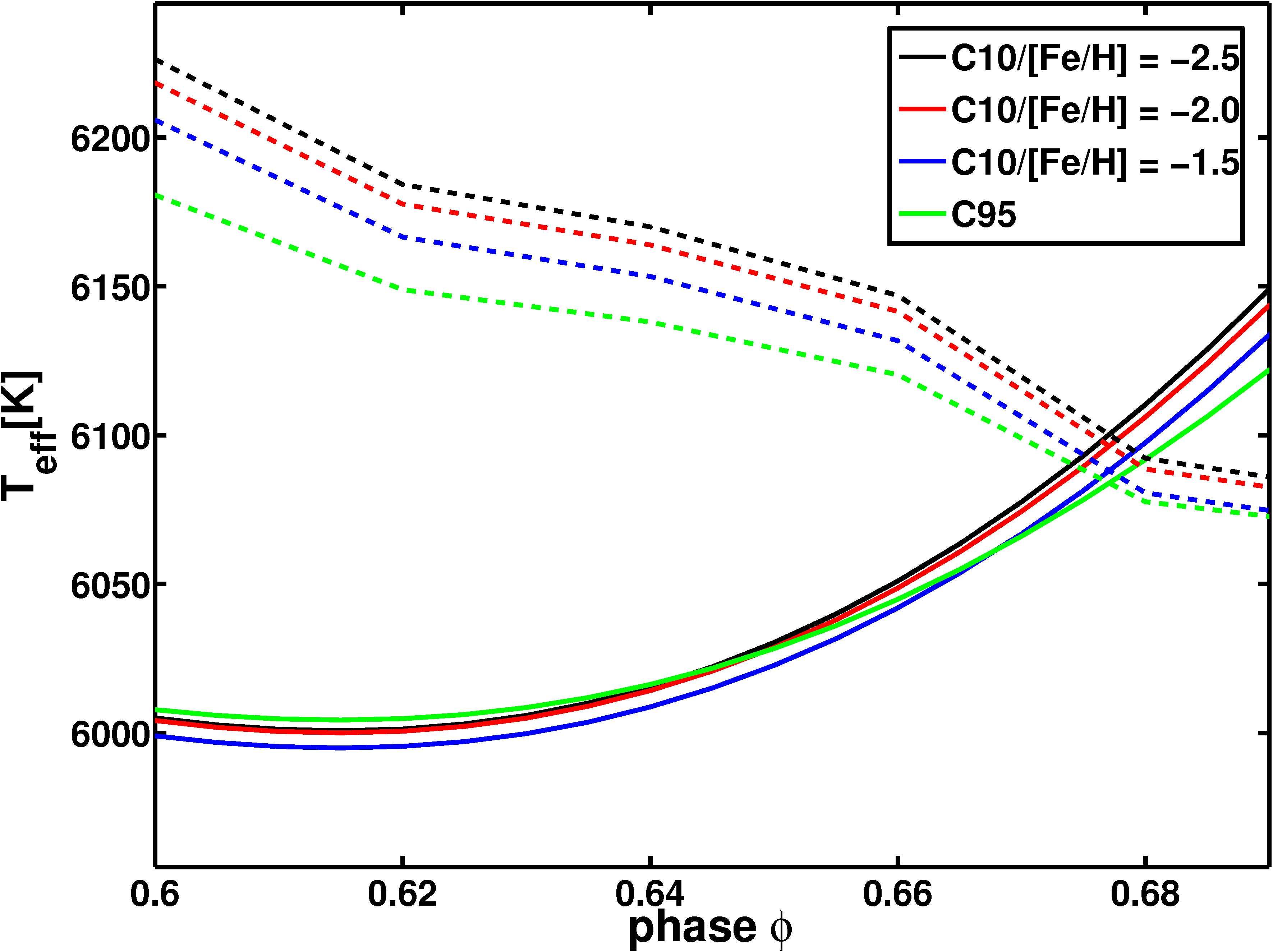} 
\caption{Using Equation\,\ref{eff_T_C95} of \citet{Clementini95} we obtain the green curves for the effective temperature of the target star RR\,15903. The black, red and blue lines indicate the temperature curves of Equation\,\ref{eff_T_C10} \citep{Casagrande10} for three different metallicities. While the solid line shows the temperature obtained by using the polynomial fit of the $V$\,band measurements, the dashed lines are computed with the template $V$\,band light curve of \citet{Layden98}. The upper panel shows the temperature for one period, while the lower panel is a blow-up of the observed phase enclosed by the two vertical black dash-dotted line in the upper panel.}
\label{RR_tempcurve}
\end{figure}

These colors are inserted into Equation\,\ref{eff_T_C95} and the resulting temperature curve over the whole period of the star is shown with the green lines in Figure\,\ref{RR_tempcurve}. For Equation\,\ref{eff_T_C10} we use the obtained colors plus a set of metallicities to calculate the effective temperature curve of the RR\,Lyrae star. We decided to use three metallicity values, $\mathrm{[Fe/H]} = -2.5$, $\mathrm{[Fe/H]} = -2.0$ and $\mathrm{[Fe/H]} = -1.5$, to calculate effective temperatures. Figure\,\ref{RR_tempcurve} illustrates that the differences introduced by the metallicity are small and therefore the temperature curve of $\mathrm{[Fe/H]} = -2.0$ is used for all stars as a temperature estimate. 

For the relations of \citet{Clementini95} and of \citet{Casagrande10}, the solid line in Figure\,\ref{RR_tempcurve} represents the temperature obtained from the polynomial fit of the $V$\,band magnitudes. The temperatures calculated with the $V$\,band template light curves of \citet{Layden98} are shown with the dotted line in Figure\,\ref{RR_tempcurve}. 

A color difference of $\Delta\mathrm{(V-I)} = 0.01$\,mag corresponds to a temperature difference of 50\,K \citep{Clementini95} in $T_{eff}$. Our $V$\,band magnitudes are not as well sampled as for the $I$ band. With individual uncertainties of $0.05$\,mag for each $V$\,band magnitude, our fit of the $V$\,band light curve becomes less certain than the $I$ band fit, where each data point has a similar uncertainty as in the $V$-band. We thus have to assume an uncertainty of at least 300\,K of the temperatures.

These temperatures are used as a first guess for spectroscopic temperature determinations with the spectral analysis program MOOG. To determine such a temperature estimate we force the abundance of the iron lines not to change with the excitation potential (EP) of the respective lines. Table\,\ref{T_eff_table} summarizes the temperatures found. 

\begin{table*}
\caption{Mean effective temperatures of the observed phase of our target stars from Equation\,\ref{eff_T_C95} and Equation\,\ref{eff_T_C10} with a metallicity of $\mathrm{[Fe/H]} = -2.0$. The first two values show the temperature obtained from using the polynomial fits of the $V$\,band. For column\,4 and column\,5 the template light curves of \citet{Layden98} are used. Moreover, the effective temperature determined by the spectral analysis program MOOG is shown in column\,6.} 
\label{T_eff_table}
\centering                          
\begin{tabular}{lcccccc}        
\hline \hline
Star & phase & $T_{\mathrm{eff/C10}}^{\mathrm{poly\,fit}}$ & $T_{\mathrm{eff/C95}}^{\mathrm{poly\,fit}}$ & $T_{\mathrm{eff/C10}}^{\mathrm{templ.\,fit}}$ & $T_{\mathrm{eff/C95}}^{\mathrm{templ.\,fit}}$ & $\mathrm{T_{eff/spec}}$ \\
 & & [K] & [K] & [K] & [K] & [K] \\
\hline
RR\,15903 & $0.60-0.69$ & 6037 & 6035 & 6143 & 6121 & 6500 \\
RR\,8645 & $0.33-0.44$ & 6529 & 6408 & 7755 & 7113 & 6500 \\
RR\,1422 & $0.52-0.62$ & 6786 & 6579 & 6703 & 6525 & 6250 \\
RR\,177 & $0.14-0.22$ & 6335 & 6269 & 6213 & 6176 & 6500 \\
RR\,11371 & $0.38-0.45$ & 6203 & 6169 & 5937 & 5950 & 6250 \\
RR\,22827 & $0.78-0.85$ & 6468 & 6360 & 5271 & 5288 & 6500 \\
RR\,122432 & $0.42-0.49$ & 7675 & 7075 & 9084 & 7649 & 6000 \\
RR\,50180 & $0.64-0.69$ & 6061 & 6053 & 6324 & 6259 & 6000 \\
RR\,143874 & $0.36-0.45$ & 6898 & 6650 & 7175 & 6812 & 6500 \\
X\,Ari & $0.47-0.55$ & 5915 & 6271 & \nodata  & \nodata & 6700 \\  
\hline
\end{tabular}
\end{table*} 

The comparison between photometric and spectroscopic temperature estimates reveals a good agreement for four of the stars, with differences smaller than 150\,K between them (Table\,\ref{T_eff_table}). For four other stars, we find deviations between 150\,K and $\sim 500$\,K, which can still be accounted for by the uncertainties due to the not very well-sampled $V$\,band light curves. In general, we find that a less good sampling of the observations in the $V$\,band introduces higher differences in the temperature estimates, while the sampling in the $I$ band is always very good. This is especially the case for the $V$\,band light curve for RR\,122432, where the temperatures deviate by more than 500\,K. Therefore the fit to the light curve is rather uncertain and the temperature estimate, which is much too high for an RR\,Lyrae star, is rejected. Thus, this star is excluded from the further comparison of photometric and spectroscopic temperatures. 

\begin{table}
\caption{The mean and median differences as well as the standard deviation between the different photometric and the spectroscopic temperature estimates $T_{\mathrm{eff/spec}}$.} 
\label{T_eff_diff_table}
\centering                          
\begin{tabular}{lrrrr}        
\hline \hline
$T^{\mathrm{fit}} - T_{\mathrm{spec}}$ & $T_{\mathrm{eff/C10}}^{\mathrm{poly\,fit}}$ & $T_{\mathrm{eff/C95}}^{\mathrm{poly\,fit}}$ & $T_{\mathrm{eff/C10}}^{\mathrm{templ.\,fit}}$ & $T_{\mathrm{eff/C95}}^{\mathrm{templ.\,fit}}$\\
 & [K] & [K] & [K] & [K] \\
\hline
mean & $39$ & $-60$ & $65$ & $95$ \\
median & $-2$ & $-87$ & $19$ & $-21$ \\
stdev $\sigma$ & $312$ & $264$ & $768$ & $597$ \\
\hline
\end{tabular}
\end{table} 

The mean, the median and the standard deviation for the differences between the photometric and the spectroscopic temperatures are shown in Table\,\ref{T_eff_diff_table}. It turns out that the polynomial fit of the $V$\,band lightcurve is preferable to the use of the templates provided by \citet{Layden98}. The templates of \citeauthor{Layden98} are calibrated on RR\,Lyrae stars with higher metallicities than the our targets. The parameters of the lightcurves change with metallicity and it is therefore not surprising that deviations from the templates are found. The relation adopted from \citet{Clementini95} leads to slightly better temperature estimates than the relation by \citet{Casagrande10}, assuming that the spectroscopic temperatures are correct. We conclude that both relations are well suited for an estimate of the temperature of an RR\,Lyrae star.

\subsection{Other stellar parameters}
\label{stellar_parameters}

The other stellar parameters, such as the surface gravity (log\,g), metallicity ([Fe/H]) and microturbulence ($\nu_t$) are determined in an iterative procedure. For the metallicity, we assume the photometric estimate as a starting point for our stellar models. The surface gravity is set to $\mathrm{log\,g} = 2.6$, a typical value for an RR\,Lyrae star at minimum light \citep[e.g.,][]{Oke66}. For the microturbulence we assume $\nu_t = 3$\,km\,s$^{-1}$.

Fe\,II is much more sensitive to the surface gravity than Fe\,I. By varying the abundance of Fe\,I and Fe\,II until the different ionization states lead to a similar abundance the value of log\,g is estimated. The microturbulence is matched such that no abundance differences with the reduced width log(EW/$\lambda$) of the Fe\,I lines are found. After each iteration a new Kurucz model is used with the updated version of the parameters and the next iteration is started, until the final set of parameters is found. Owing to the low S/N for two stars the difference between Fe\,I and Fe\,II can only be reduced by assuming unphysical stellar parameters. Therefore we allow for differences of 0.2\,dex for the LMC targets and 0.3\,dex for the SMC RR\,Lyrae stars. Furthermore we test our log\,g values using photometrically derived gravities and find good agreement between both methods. The final set of stellar parameters is shown in Table\,\ref{stellar_parameter}. 

\begin{table}
\caption{The final set of parameters from the EW abundance determination.} 
\label{stellar_parameter}
\centering                          
\begin{tabular}{lcccccc}        
\hline \hline
Star & $T_{\mathrm{eff}}$ & log\,g & [Fe/H] & $\nu_t$ \\
 & [K] &  & [dex] & [km/s$^{-1}$] \\
\hline
HD74000 & 6100 & 3.5 & $-2.08$ & 1.5 \\
X\,Ari & 6700 & 2.1 & $-2.61$ & 2.6 \\  
G64-12 & 6350 & 4.4 & $-3.13$ & 1.5 \\
\hline
RR\,15903 & 6500 & 2.6 & $-2.24$ & 3.5 \\
RR\,8645 & 6500 & 3.0 & $-1.97$ & 5.0 \\
RR\,1422 & 6250 & 2.0 & $-2.49$ & 4.0 \\
RR\,177 & 6500 & 2.6 & $-2.67$ & 3.8 \\
RR\,11371 & 6250 & 2.6 & $-2.18$ & 3.2 \\
RR\,22827 & 6500 & 1.5 & $-2.49$ & 2.5 \\
RR\,122432 & 6000 & 2.6 & $-2.36$ & 4.5 \\
RR\,50180 & 6000 & 2.0 & $-2.36$ & 4.5 \\
RR\,143874 & 6500 & 2.0 & $-2.34$ & 2.0 \\
\hline
\end{tabular}
\end{table}

\subsection{Comparison of stellar parameters for the standard stars with literature values}

For HD\,74000, we found data on the stellar parameters in \citet{Peterson78}, \citet{Beveridge94}, and \citet{Fulbright00}. Their estimates agree very well with ours. Only the log\,g value is higher by about 0.6\,dex (see Table\,\ref{stellar_parameter_standards_literature}). 

For X\,Ari, we estimate slightly higher temperature and lower log\,g values than found by \citet{Clementini95} and \citet{Lambert96} (see Table\,\ref{stellar_parameter_standards_literature}). However, this is an effect of having observed X\,Ari at a different phases. While we have observed X\,Ari at $\Phi = 0.5$, \citet{Lambert96} have taken their data at $\Phi = 0.38$ and \citet{Clementini95} at $\Phi = 0.75$. The metallicity is in very good agreement with the literature, while the microturbulence is a bit lower than the literature values. 

For G64-12, the agreement of our estimates with the values by \citet{Boesgaard06} is very good for all the parameters.

\begin{table*}
\caption{Stellar parameters of standard stars from the literature.} 
\label{stellar_parameter_standards_literature}
\centering                          
\begin{tabular}{lcccccc}        
\hline \hline
Star & $T_{\mathrm{eff}}$ & log\,g & [Fe/H] & $\nu_t$ & Source\\
 & [K] &  & [dex] & [km/s$^{-1}$] & \\
\hline
HD74000 & 6250 & 4.50 & $-1.80$ & \nodata & \citet{Peterson78} \\
HD74000 & 6090 & 4.15 & $-2.07$ & 1.30 & \citet{Beveridge94} \\
HD74000 & 6025 & 4.10 & $-2.00$ & 1.20 & \citet{Fulbright00} \\
X\,Ari & 6100 & 2.60 & $-2.5$ & 3.0 & \citet{Lambert96}$_{\mathrm{phot}}$ \\
X\,Ari & 6500 & 2.70 & $-2.5$ & 4.4 & \citet{Lambert96} \\
X\,Ari & 6109 & 2.60 & $-2.5$ & 4.5 & \citet{Clementini95} \\
G64-12 & 6350 & 4.40 & $-3.0$ & 1.5 & \citet{Boesgaard06} \\
\hline
\end{tabular}
\tablecomments{phot - estimates from photometric data, otherwise values from spectroscopic data} 
\end{table*}

\subsection{Systematic parameter uncertainties}

We test the sensitivity of the abundances to the stellar parameters by altering one single parameter of X\,Ari at a time and calculating the abundance difference (Table\,\ref{list_error_abundance_EW}). For the temperature we assume an uncertainty of $\Delta T = 300$\,K, which corresponds to the uncertainty of our photometric temperature estimate. For the surface gravity we adopt an uncertainty of $\Delta\mathrm{log\,g} = 0.5$\,dex, while for the microturbulence $\Delta \mathrm{\nu_t} = 0.5$\,km/s$^{-1}$ is assumed. These differences are adopted as uncertainties for all RR\,Lyrae stars in our sample.

\begin{table}
\caption{Abundance differences due to parameter uncertainties, applied to X\,Ari as a proxy for all target stars} 
\label{list_error_abundance_EW}
\centering                          
\begin{tabular}{lrrr}        
\hline \hline
Element & $\Delta T = 300$\,K & $\Delta\mathrm{log\,g} = 0.5$ & $\Delta \mathrm{\nu_t} = 0.5$\,km/s$^{-1}$ \\
\hline
Na\,I & $-0.21$ & 0.02 & 0.05 \\
Mg\,I & $-0.16$ & 0.02 & 0.10 \\  
Al\,I & $-0.24$ & 0.02 & 0.01 \\
Si\,I & $-0.24$ & 0.03 & 0.04 \\
Ca\,I & $-0.22$ & 0.02 & 0.08 \\
Sc\,II & $-0.15$ & $-0.16$ & 0.07 \\
Ti\,I & $-0.15$ & 0.03 & 0.02 \\
Ti\,II & $-0.18$ & $-0.12$ & 0.13 \\
Cr\,I & $-0.28$ & 0.02 & 0.01 \\
Cr\,II & $-0.30$ & $-0.01$ & 0.19 \\
Mn\,I & $-0.30$ & 0.03 & 0.01 \\
Fe\,I & $-0.28$ & 0.03 & 0.08 \\
Fe\,II & $-0.06$ & $-0.16$ & 0.08 \\
Co\,I & $-0.29$ & 0.02 & 0.02 \\
Ni\,I & $-0.51$ & 0.17 & 0.15 \\
Sr\,II & $-0.21$ & $-0.13$ & 0.12 \\
Ba\,II & $-0.26$ & $-0.08$ & 0.01 \\
\hline
\end{tabular}
\end{table} 

For neutral species the abundance ratios show the largest dependence on the estimated T$_{\mathrm{eff}}$. These elements are rather insensitive to changes in the surface gravity or microturbulence. The surface gravity plays a much larger role for the ionized elements.


\section{Results: abundance ratios}
\label{Abundances}

We use two independent approaches to determine the abundances of the nine target and three standard stars. First, we calculate the the abundances directly from the EWs using MOOG's \textit{abfind} driver. For the second approach, we synthesize the spectral features of the stars. Both approaches are performed with the 2010 version of the spectral analysis program MOOG \citep{Sneden73}.

We start with the abundance determination from the EW measurements. Model atmosphere parameters are obtained through an iterative process (see Section\,\ref{Parameter}). These best model parameters are used to determine the abundances of up to 14 different elements (Na\,I, Mg\,I, Al\,I, Si\,I, Ca\,I, Sc\,II, Ti, Cr, Mn\,I, Fe, Co\,I, Ni\,I, Sr\,II, Ba\,II). For three of the elements (Ti, Cr, Fe) we are able to determine the abundances both for neutral and singly ionized line transitions. 

\setlength{\tabcolsep}{3.7pt}
\begin{table*}
\caption{Abundances from EW measurements and spectral synthesis of the lines.} 
\label{abundances_table}
\centering                          
\begin{tabular}{lrccrccrccrccrccrcc}        
\hline \hline
star & \multicolumn{3}{c}{[Fe/H]} & \multicolumn{3}{c}{[Fe\,I/H]} & \multicolumn{3}{c}{[Fe\,II/H]}  \\
 & synth & $\sigma$ & $N$ & EW & $\sigma$ & $N$ & EW & $\sigma$ & $N$ \\
\hline
HD74000 & $-2.15$ & $0.15$ & $10$ & $-2.08$ & $0.28$ & $28$ & $-2.12$ & $0.31$ & $6$ \\
X\,Ari & $-2.58$ & $0.10$ & $9$ & $-2.61$ & $0.23$ & $26$ & $-2.62$ & $0.28$ & $5$ \\
G64-12 & $-3.06$ & $0.20$ & $6$ & $-3.13$ & $0.28$ & $20$ & $-3.02$ & $0.36$ & $2$ \\
\hline
RR\,15903 & $-2.34$ & $0.00$ & $2$ & $-2.24$ & $0.26$ & $11$ & $-2.32$ & $0.24$ & $3$ \\
RR\,8645 & $-1.72$ & $0.33$ & $4$ & $-1.97$ & $0.26$ & $8$ & $-1.94$ & $0.41$ & $3$ \\
RR\,1422 & $-2.26$ & $0.17$ & $4$ & $-2.49$ & $0.32$ & $13$ & $-2.23$ & $0.31$ & $4$ \\
RR\,177 & $-2.57$ & \nodata & $1$ & $-2.67$ & $0.33$ & $10$ & $-2.50$ & $0.48$ & $3$ \\
RR\,11371 & $-2.10$ & \nodata & $1$ & $-2.18$ & $0.24$ & $6$ & $-2.16$ & $0.48$ & $2$ \\
RR\,22827 & $-2.42$ & $0.11$ & $4$ & $-2.50$ & $0.24$ & $14$ & $-2.63$ & $0.08$ & $5$ \\
RR\,122432 & \nodata & \nodata & \nodata & $-2.36$ & $0.12$ & $9$ & $-2.31$ & $0.47$ & $3$ \\
RR\,50180 & \nodata & \nodata & \nodata & $-2.36$ & $0.33$ & $13$ & $-2.02$ & $0.34$ & $2$ \\
RR\,143874 & \nodata & \nodata & \nodata & $-2.34$ & $0.17$ & $8$ & $-2.34$ & $0.39$ & $3$ \\
\hline \hline
star & \multicolumn{6}{c}{[Na\,I/Fe\,I]} & \multicolumn{6}{c}{[Mg\,I/Fe\,I]} & \multicolumn{6}{c}{[Al\,I/Fe\,I]}  \\
 & synth & $\sigma$ & $N$ & EW & $\sigma$ & $N$ & synth & $\sigma$ & $N$ & EW & $\sigma$ & $N$ & synth & $\sigma$ & $N$ & EW & $\sigma$ & $N$ \\
\hline
HD74000 & $0.70$ & \nodata & $1$ & $0.57$ & $0.23$ & $2$ & $0.60$ & $0.00$ & $2$ & $0.68$ & $0.71$ & $4$ & $0.50$ & \nodata & $1$ & \nodata & \nodata & \nodata \\
X\,Ari & $-0.05$ & \nodata & $1$ & $0.17$ & $0.24$ & $2$ & $0.35$ & $0.10$ & $2$ & \nodata & \nodata & \nodata & $-0.50$ & \nodata & $1$ & $-0.65$ & $0.37$ & $2$ \\
G64-12 & $-0.40$ & \nodata & $1$ & $-0.30$ & $0.38$ & $2$ & $-0.03$ & $0.06$ & $2$ & $0.14$ & \nodata & $1$ & $-0.40$ & \nodata & $1$ & \nodata & \nodata & \nodata \\
\hline
RR\,15903 & $0.10$ & \nodata & $1$ & $0.02$ & \nodata & $1$ & $0.55$ & $0.10$ & $2$ & $0.54$ & $0.34$ & $2$ & \nodata & \nodata & \nodata & \nodata & \nodata & \nodata \\
RR\,8645 & $0.10$ & \nodata & $1$ & $-0.05$ & \nodata & $1$ & $0.75$ & $0.49$ & $2$ & $0.93$ & \nodata & $1$ & \nodata & \nodata & \nodata & \nodata & \nodata & \nodata \\
RR\,1422 & \nodata & \nodata & \nodata & \nodata & \nodata & \nodata & $0.50$ & $0.00$ & $2$ & \nodata & \nodata & \nodata & $0.50$ & \nodata & $1$ & \nodata & \nodata & \nodata \\
RR\,177 & \nodata & \nodata & \nodata & \nodata & \nodata & \nodata & $-0.50$ & \nodata & $1$ & \nodata & \nodata & \nodata & \nodata & \nodata & \nodata & \nodata & \nodata & \nodata \\
RR\,11371 & \nodata & \nodata & \nodata & \nodata & \nodata & \nodata & \nodata & \nodata & \nodata & \nodata & \nodata & \nodata & \nodata & \nodata & \nodata & \nodata & \nodata & \nodata \\
RR\,22827 & \nodata & \nodata & \nodata & \nodata & \nodata & \nodata & $0.35$ & $0.10$ & $2$ & \nodata & \nodata & \nodata & $0.40$ & \nodata & $1$ & \nodata & \nodata & \nodata \\
RR\,122432 & \nodata & \nodata & \nodata & \nodata & \nodata & \nodata & \nodata & \nodata & \nodata & \nodata & \nodata & \nodata & \nodata & \nodata & \nodata & \nodata & \nodata & \nodata \\
RR\,50180 & \nodata & \nodata & \nodata & \nodata & \nodata & \nodata & $0.10$ & \nodata & $1$ & $0.03$ & $0.35$ & $2$ & \nodata & \nodata & \nodata & \nodata & \nodata & \nodata \\
RR\,143874 & \nodata & \nodata & \nodata & \nodata & \nodata & \nodata & \nodata & \nodata & \nodata & \nodata & \nodata & \nodata & \nodata & \nodata & \nodata & \nodata & \nodata & \nodata \\
\hline \hline
star & \multicolumn{6}{c}{[Ca\,I/Fe\,I]} & \multicolumn{6}{c}{[Sc\,II/Fe\,II]} & \multicolumn{6}{c}{[Ti\,II/Fe\,II]} \\
 & synth & $\sigma$ & $N$ & EW & $\sigma$ & $N$ & synth & $\sigma$ & $N$ & EW & $\sigma$ & $N$ & synth & $\sigma$ & $N$ & EW & $\sigma$ & $N$ \\
\hline
HD74000 & $0.20$ & $0.40$ & $5$ & $0.24$ & \nodata & $1$ & $-0.05$ & \nodata & $1$ & $-0.04$ & \nodata & $1$ & $0.19$ & $0.18$ & $8$ & $0.52$ & $0.37$ & $12$\\
X\,Ari & $0.33$ & $0.22$ & $4$ & $0.50$ & $0.51$ & $3$ & $0.50$ & \nodata & $1$ & $0.36$ & \nodata & $1$ & $0.16$ & $0.23$ & $6$ & $0.70$ & $0.24$ & $14$\\
G64-12 & $0.50$ & $0.00$ & $2$ & $0.70$ & $0.59$ & $2$ & $0.10$ & \nodata & $1$ & $0.19$ & \nodata & $1$ & $0.10$ & $0.21$ & $3$ & $0.44$ & $0.27$ & $8$\\
\hline
RR\,15903 & $0.75$ & $0.49$ & $2$ & \nodata & \nodata & \nodata & \nodata & \nodata & \nodata & \nodata & \nodata & \nodata & $0.10$ & $0.39$ & $2$ & $0.24$ & $0.27$ & $6$\\
RR\,8645 & $0.00$ & \nodata & $1$ & \nodata & \nodata & \nodata & \nodata & \nodata & \nodata & \nodata & \nodata & \nodata & $0.33$ & $0.15$ & $4$ & $0.38$ & \nodata & $1$\\
RR\,1422 & $0.50$ & \nodata & $1$ & \nodata & \nodata & \nodata & \nodata & \nodata & \nodata & \nodata & \nodata & \nodata & $0.33$ & $0.43$ & $3$ & $0.24$ & $0.35$ & $3$\\
RR\,177 & \nodata & \nodata & \nodata & \nodata & \nodata & \nodata & $0.30$ & \nodata & $1$ & $0.34$ & \nodata & $1$ & $0.00$ & \nodata & $1$ & $-0.20$ & $0.34$ & $4$\\
RR\,11371 & $0.33$ & $0.06$ & $2$ & \nodata & \nodata & \nodata & \nodata & \nodata & \nodata & \nodata & \nodata & \nodata & \nodata & \nodata & \nodata & $0.26$ & $0.46$ & $3$\\
RR\,22827 & $0.35$ & $0.10$ & $2$ & $0.38$ & $0.23$ & $3$ & $-0.20$ & \nodata & $1$ & \nodata & \nodata & \nodata & $0.20$ & \nodata & $1$ & $0.00$ & $0.45$ & $4$\\
RR\,122432 & \nodata & \nodata & \nodata & \nodata & \nodata & \nodata & \nodata & \nodata & \nodata & \nodata & \nodata & \nodata & \nodata & \nodata & \nodata & $-0.66$ & $0.24$ & $4$\\
RR\,50180 & \nodata & \nodata & \nodata & \nodata & \nodata & \nodata & \nodata & \nodata & \nodata & \nodata & \nodata & \nodata & $0.00$ & \nodata & $1$ & $-0.01$ & $0.63$ & $2$\\
RR\,143874 & $0.20$ & \nodata & $1$ & \nodata & \nodata & \nodata & \nodata & \nodata & \nodata & \nodata & \nodata & \nodata & \nodata & \nodata & \nodata & $0.05$ & $0.24$ & $4$\\
\hline \hline
star & \multicolumn{6}{c}{[Cr\,I/Fe\,I]} & \multicolumn{6}{c}{[Sr\,II/Fe\,II]} & \multicolumn{6}{c}{[Ba\,II/Fe\,II]}  \\
 & synth & $\sigma$ & $N$ & EW & $\sigma$ & $N$ & synth & $\sigma$ & $N$ & EW & $\sigma$ & $N$ & synth & $\sigma$ & $N$ & EW & $\sigma$ & $N$ \\
\hline
HD74000 & $-0.20$ & $0.39$ & $2$ & $-0.18$ & $0.19$ & $4$ & $0.60$ & \nodata & $2$ & $0.39$ & \nodata & $1$ & $0.20$ & $0.59$ & $2$ & $0.49$ & $0.22$ & $3$\\
X\,Ari & $0.05$ & $0.39$ & $2$ & $-0.11$ & $0.38$ & $3$ & $-0.40$ & $0.20$ & $2$ & $-0.39$ & \nodata & $1$ & $-0.30$ & \nodata & $1$ & $-0.12$ & $0.41$ & $2$\\
G64-12 & $0.00$ & \nodata & $1$ & \nodata & \nodata & \nodata & \nodata & \nodata & \nodata & \nodata & \nodata & \nodata & \nodata & \nodata & \nodata & $0.81$ & $0.35$ & $2$\\
\hline
RR\,15903 & \nodata & \nodata & \nodata & \nodata & \nodata & \nodata & $0.40$ & \nodata & $1$ & \nodata & \nodata & \nodata & $0.20$ & \nodata & $1$ & $0.08$ & $0.42$ & $4$ \\
RR\,8645 & \nodata & \nodata & \nodata & \nodata & \nodata & \nodata & \nodata & \nodata & \nodata & \nodata & \nodata & \nodata & $0.50$ & $0.00$ & $2$ & $0.30$ & $0.16$ & $2$ \\
RR\,1422 & $0.10$ & $0.20$ & $2$ & $-0.24$ & $0.03$ & $2$ & $0.25$ & $0.49$ & $2$ & \nodata & \nodata & \nodata & \nodata & \nodata & \nodata & $0.04$ & $0.17$ & $4$ \\
RR\,177 & $0.30$ & \nodata & $1$ & $0.02$ & $0.02$ & $2$ & $-0.40$ & \nodata & $1$ & \nodata & \nodata & \nodata & \nodata & \nodata & \nodata & $0.32$ & $0.83$ & $2$ \\
RR\,11371 & \nodata & \nodata & \nodata & \nodata & \nodata & \nodata & \nodata & \nodata & \nodata & \nodata & \nodata & \nodata & \nodata & \nodata & \nodata & $-0.50$ & $0.37$ & $2$ \\
RR\,22827 & $0.50$ & \nodata & $1$ & $0.38$ & \nodata & $1$ & $0.10$ & $0.20$ & $2$ & \nodata & \nodata & \nodata & \nodata & \nodata & \nodata & $0.35$ & $0.89$ & $2$ \\
RR\,122432 & \nodata & \nodata & \nodata & \nodata & \nodata & \nodata & \nodata & \nodata & \nodata & \nodata & \nodata & \nodata & \nodata & \nodata & \nodata & $-0.61$ & $1.34$ & $2$ \\
RR\,50180 & \nodata & \nodata & \nodata & \nodata & \nodata & \nodata & \nodata & \nodata & \nodata & \nodata & \nodata & \nodata & \nodata & \nodata & \nodata & $-0.20$ & $0.54$ & $4$ \\
RR\,143874 & \nodata & \nodata & \nodata & \nodata & \nodata & \nodata & \nodata & \nodata & \nodata & \nodata & \nodata & \nodata & \nodata & \nodata & \nodata & $0.15$ & $0.85$ & $3$ \\
\hline
\end{tabular}
\end{table*} 

The spectral features that are significantly larger than the background noise are identified. These lines are further investigated by spectral synthesis. All features close to the line of interest are fitted by the spectral synthesis model to determine abundance values for all features present and measureable. In this way we properly account for blended lines, to yield the correct abundance of the line of interest. The resulting mean abundances for [Fe/H] and [X/Fe] of the synthesis for the different elements are shown in Table\,\ref{abundances_table}. Here, X stands for all investigated elements except Fe. 

For each species X, we have computed the straight mean and standard deviation for the abundance ratios [X/Fe], as described in Section\,\ref{Uncertainties}. These are listed in Table\,\ref{abundances_table}, together with the number of lines, N, used to determine the abundances.

For those elements with a larger number of measurable lines (e.g., Fe and Ti), the random error, $\sigma/\sqrt{N}$, is small and the total error on [X/Fe] for those species is dominated by the systematic uncertainty. Elements that only have one line accessible for determining the abundance ratio were assigned a representative, minimum random error of 0.2\,dex, a typical value based on the S/N characteristics and continuum setting for the respective lines in comparable abundance studies. The total error bar shown in the following Figures includes the systematic and statistical contributions summed in quadrature.

Due to the low S/N level of the target stars, we inspect the profiles of the features very carefully. The knowledge of the expected line profile of the feature leads to a more conservative abundance estimate when using the spectral synthesis as opposed to the EW method. While using the EW on low S/N data, it may sometimes be difficult to distinguish between real features and additional noise. If the abundances from the EW measurements agree within the uncertainties with the results obtained from the spectral synthesis we show them in Table\,\ref{abundances_table} as well. The standard deviations $\sigma$ of the EW abundances are computed in the same way as described for the spectral synthesis. For elements with only a few available lines, the spectral synthesis abundances are anyhow always taken as more reliable.

Our target stars have very similar stellar parameters as the RR\,Lyrae stars investigated in \citet{Hansen11}. Therefore, we adopt the same NLTE corrections, which are larger than our uncertainties, as they did. For aluminum a correction of $\Delta_{\mathrm{NLTE}} = +0.7$\,dex \citep{Andrievsky08} is used. The NLTE correction for strontium is adopted from \citet{Mashonkina01} with a value of $\Delta_{\mathrm{NLTE}} = +0.6$\,dex.

\subsection{Iron}

For the iron content of the stars, we adopt the data from the EWs as well as from the spectral synthesis. We are able to measure EWs of eight to 34 reliable iron lines per star. For the other elements we consider only the synthesis abundances since there are many fewer lines.

The extremely low photometric metallicity of some of the targets cannot be confirmed spectroscopically. For our target stars we find spectroscopic metallicities between $-2.67\,\mathrm{dex} \leq \mathrm{[Fe/H]} \leq -1.97$\,dex (see Table\,\ref{abundances_table}). The RR\,Lyrae star with $\mathrm{[Fe/H]} = -2.67$\,dex is the most metal-poor star found so far in the MCs (see Figure\,\ref{spectra_Mg_LMC}). 

Measuring the EW, we detect for all target stars several Fe\,I lines and at least two reliable Fe\,II lines. By construction the abundances of both ionisation states should be the same, but due to the low S/N of the data we allow for small differences to obtain reasonable stellar parameters (see Section\,\ref{stellar_parameters}). Therefore, a small difference of on average $-0.07 \pm 0.15$\,dex is found between the abundances of Fe\,I and Fe\,II from EW measurements for our target stars. 

Some of the iron lines are weak and the synthesizing process did not provide a significant result within our uncertainty margins. However, for the lines that are synthesizable, very good agreement with the abundances from the EW measurements is obtained. We are therefore confident that the values obtained from the EW are reliable owing to the large number of lines used. 

The low S/N is responsible that the synthesis of the iron lines in our SMC targets did not yield any results within our error margins of 0.3\,dex. We therefore rely on the EW measurements for the iron content of these stars. The individual measurements of the lines have large uncertainties, but we were able to measure a large number of lines, which leads to a reliable determination of the abundance.

\subsection{$\alpha$-elements}
\label{alpha}

We now consider the abundance ratios of the $\alpha$-elements magnesium, calcium and titanium. We note that we cannot synthesize any silicon line to a satisfying accuracy, because they are in the blue part of the spectrum where the low S/N makes abundance determinations particularly difficult. 

Synthesis abundances for $\alpha$-elements are derived for five LMC and one SMC RR\,Lyrae star (Table\,\ref{abundances_table}). For four of these LMC stars we find a mean value of $\mathrm{[Mg/Fe]} = 0.54 \pm 0.19$\,dex.  For our SMC RR\,Lyrae target RR\,50180 a lower abundance value of $\mathrm{[Mg/Fe]} = 0.10$\,dex is obtained. For RR\,177 in the LMC we obtain a quite low value of $\mathrm{[Mg/Fe]} = -0.50$\,dex. We discuss this special case in detail below. The mean abundances for calcium and titanium are slightly lower. We find $\mathrm{[Ca/Fe]} = 0.39 \pm 0.30$\,dex and $\mathrm{[Ti/Fe]} = 0.19 \pm 0.16$\,dex for the LMC RR\,Lyrae stars. For the SMC star RR\,143874 a value of $\mathrm{[Ca/Fe]} = 0.20$\,dex is obtained, while for RR\,50180 $\mathrm{[Ti/Fe]} = 0.00$\,dex is found. We caution that the S/N of the SMC targets is considerably lower than that of the LMC objects.

\begin{figure}
\includegraphics[width=0.47\textwidth]{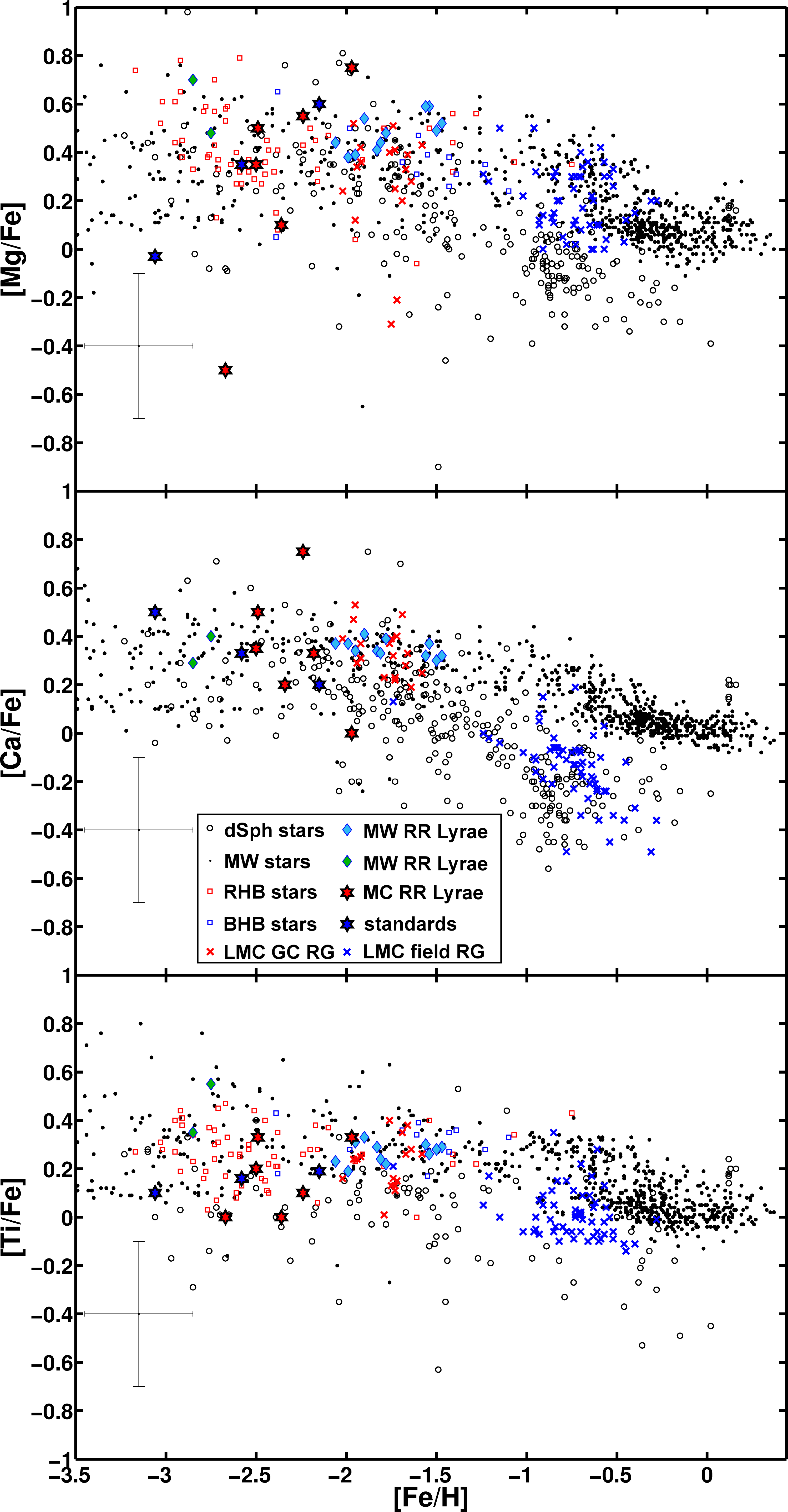} 
\caption{Synthesized abundances of the light $\alpha$-elements Mg, Ca and Ti. We find good agreement with the abundances obtained for other stars at similar metallicity. Especially the MW RR\,Lyrae stars from \citet[cyan diamonds]{For11} and \citet[green diamonds]{Hansen11} show good agreement with our sample of RR\,Lyrae stars. Additionally we find good agreement with the BHB (blue squares) and RHB stars (red squares) of \citet{For10, Preston06}, as well as with LMC GC RGs (red crosses) from \citet{Mucciarelli10}. For the LMC RR\,Lyrae stars a mean overabundance of [$\alpha$/Fe] = 0.36\,dex is found. }
\label{alpha_Fe}
\end{figure}

A simple mean of all measurements of the $\alpha$ abundances of the LMC RR\,Lyrae stars, excluding RR\,177, leads to an $\alpha$-enhancement of $\mathrm{[\alpha/Fe]} = 0.36 \pm 0.25$\,dex for the LMC stars. Due to the small number of measurements and the low S/N level of the SMC spectra a meaningful mean value for the SMC RR\,Lyrae stars cannot be computed. The tendency of lower $\alpha$-enhancement seen from the three measurements is not significant within the uncertainties. 

The LMC abundances are in very good agreement with the literature values for the MW RR\,Lyrae stars \citep{For11, Hansen11} as well as with the HB stars of \citet{For10} and \citet{Preston06}. \citet{Johnson06} spectroscopically investigated red giants in four LMC GCs with metallicities between $-2.2 \leq \mathrm{[Fe/H]} \leq -1.2$\,dex, while in \citet{Mucciarelli09} the metallicity of red giants in three LMC GCs with was measured to be $-1.95 \leq \mathrm{[Fe/H]} \leq -1.65$\,dex. The stars in these clusters show generally stronger $\alpha$-enhancements with lower metallicities, but the $\alpha$-enhancement of the GCs studied by \citet{Johnson06} is less pronounced than in the GCs studied by \citet{Mucciarelli09}. The mean Mg-overabundance for the most metal-poor stars of \citet{Mucciarelli09} is between $0.4 \leq \mathrm{[Mg/Fe]} \leq 0.5$\,dex, in very good agreement with the RR\,Lyrae stars. \citet{Mucciarelli10} obtained the abundances of calcium and titanium for these clusters. The overabundance of these elements is lower than for [Mg/Fe], similarly to our results. In \citet{Johnson06} the $\alpha$-enhancement is in general lower. The mean Mg-overabundance for the most metal-poor cluster is 0.2\,dex. For the other $\alpha$-elements, no or only slight enrichment was found for the metal-poor cluster stars.

For the metal-poor stars located in the MW halo and in the Galactic dSph galaxies a general trend of $\alpha$-enhancement has been found due to preferential early enrichment by supernovae of type\,II. The spread of abundances between single stars is quite large, which may be due to localized, inhomogeneous enrichment. However, the differences in abundance are in general much larger than for stars with metallicity $\mathrm{[Fe/H]} \geq -1$\,dex, which are mainly located in the MW disk. Our investigated stars of the MCs fit quite well into the picture of $\alpha$-enhancement, even though the actual abundances determined by us for the individual stars differ by several tenths of dex. As for the other galaxies that may indicate localized inhomogeneous enrichment \citep[see, e.g.,][]{Marcolini08} Therefore no claim of RR\,Lyrae stars in the MCs being different from the stars in the MW halo and in the dSphs can be made and we suggest that the MCs underwent a similar early chemical evolution.

For metal-poor stars it has been known for a long time \citep[e.g.,][]{Wallerstein63} that they are enhanced in $\alpha$-elements, typically by [$\alpha/\mathrm{Fe]} \sim 0.4$\,dex. In Figure\,\ref{alpha_Fe}, we compare the abundances of our target stars with literature values for different samples. The black dots represent stars in the MW thin and thick disk as well as from the halo \citep[][and references therein]{Bonifacio09, Cayrel04, Lai08, Venn04}. The black open circles show the abundance of red giants from different MW dSph companions, namely Bootes\,I \citep{Feltzing09, Norris10c}, Carina \citep{Koch08a, Venn12}, Coma Berenices \citep{Frebel10b}, Draco \citep{Cohen09} Fornax \citep{Letarte10}, Hercules \citep{Aden11, Koch08b}, Leo\,II \citep{Shetrone09}, Sagittarius \citep{Monaco05, Sbordone07}, Sculptor \citep{Frebel10a}, Sextans \citep{Aoki09}, Ursa Major\,II \citep{Frebel10b} and Ursa Minor \citep{Sadakane04}. RR\,Lyrae stars are on the horizontal branch and therefore we show the abundances of red horizontal branch (RHB) stars from \citet{For10} and \citet{Preston06} as well as the blue horizontal branch (BHB) stars from \citet{For10} with red and blue squares, respectively. In \citet{For11} eleven nearby Galactic field RR\,Lyrae stars were spectroscopically investigated. The analysis revealed a very similar elemental abundance for all RR\,Lyrae stars, shown with cyan diamonds, with a mean overabundance of $\mathrm{[Ca/Fe]} = 0.34$\,dex and $\mathrm{[Mg/Fe]} = 0.48$\,dex. In \citet{Hansen11} two very metal-poor Galactic RR\,Lyrae stars were investigated. The $\alpha$-enhancement for these two stars, illustrated by green diamonds, is similar to the results found by \citet{For11}. 

The very metal-poor field population of the MCs has not yet been investigated. Instead we include abundances of red giants from the LMC field, which sample primarily the intermediate-age population, \citep[blue crosses]{Pompeia08} and LMC GCs \citep[red crosses]{Johnson06, Mucciarelli09, Mucciarelli10} in our comparison.

\subsection{Light odd-Z elements}

The light odd-Z elements are investigated via the resonance D-line doublet of sodium at 5889.9\,\AA \ and 5895.9\,\AA, as well as through the aluminum lines at 3944\,\AA \ and 3961\,\AA. For the sodium lines, the spectrum is carefully inspected for interstellar contamination. Fortunately, the radial velocity is high enough to avoid major contamination by interstellar lines. Regrettably we can obtain synthesized abundances for these two elements with acceptable uncertainties for two stars only (Table\,\ref{abundances_table}). 

\begin{figure}
\includegraphics[width=0.47\textwidth]{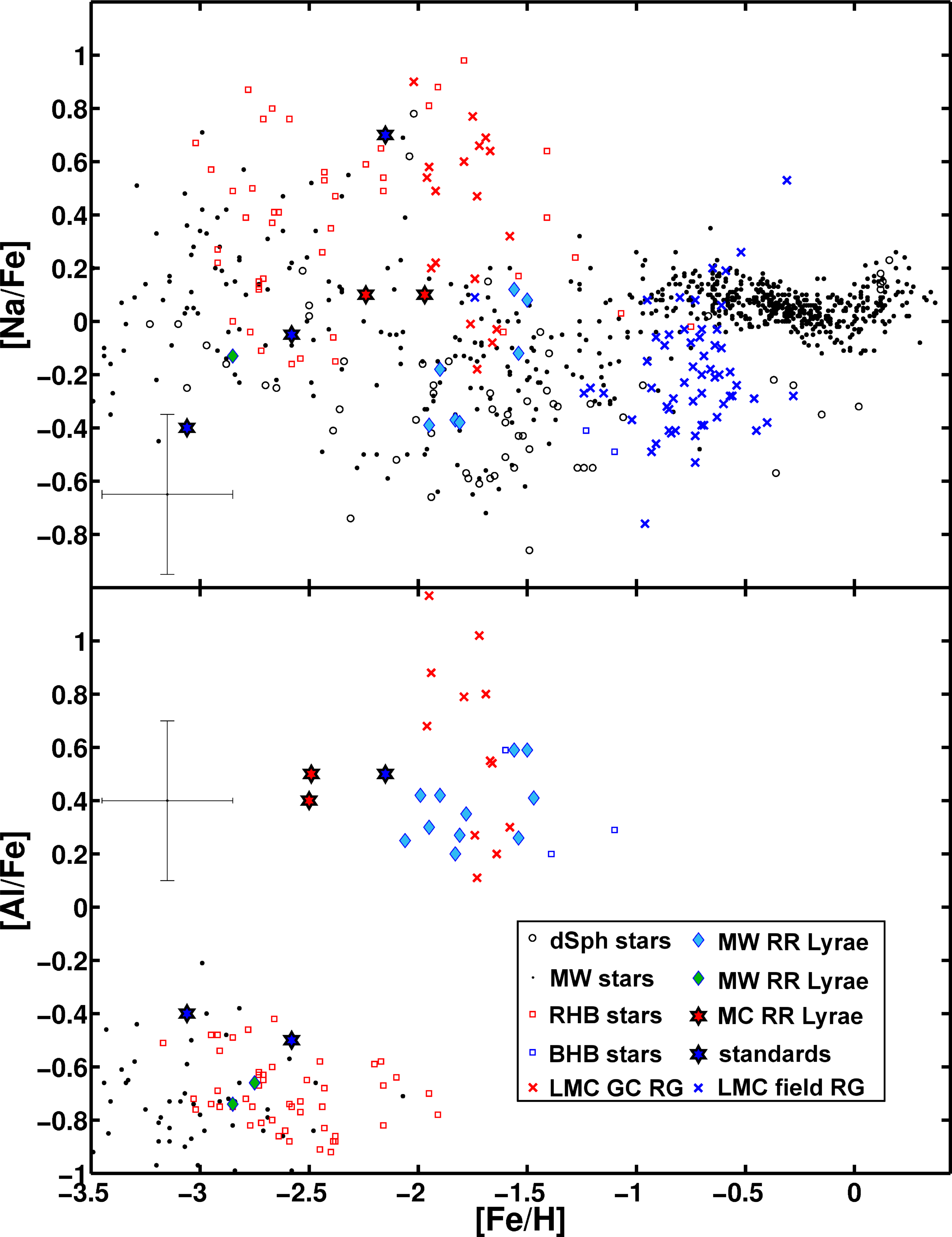} 
\caption{Abundance ratio of sodium to iron (upper panel) and aluminum to iron (lower panel) vs. [Fe/H]. The coding of the different types of stars is the same as in Figure\,\ref{alpha_Fe}. We see that the LMC RR\,Lyrae stars follow the trend of sodium enhancement with lower iron abundance. For the aluminum abundance a bimodal distribution is found. The RR\,Lyrae stars with measureable aluminum abundance belong to the group of aluminum enhanced stars, even though they are more metal-poor than the other stars of this group. }
\label{light_odd_Z}
\end{figure}

In Figure\,\ref{light_odd_Z} the same literature sources are used as in Figure\,\ref{alpha_Fe} and described in Section\,\ref{alpha}. In the low metallicity regime of our target stars the sodium abundances of the MW and the dSph stars are very similar. The two LMC RR\,Lyrae stars are slightly enhanced in their sodium abundance and are in good agreement with those stars of the MW and the dSphs. The stars in the old metal-poor GCs of the LMC have very similar sodium abundances as our RR\,Lyrae targets. \citet{Johnson06} found solar mean ratios for their LMC clusters, while \citet{Mucciarelli10} obtained sodium enhancement for most of the stars in the GCs. 

A bimodal distribution of the aluminum abundances is observed investigating BHB, RHB and RR\,Lyrae stars in the MW. While the more metal-rich RR\,Lyrae and the BHB stars are enhanced in aluminum, the metal-poor RR\,Lyrae and the RHB stars show an underabundance in [Al/Fe]. There is no explanation for this phenomenon yet. However, we find that our two target stars have aluminum abundances similar to the Galactic RR\,Lyrae and the BHB stars, even though the LMC RR\,Lyrae stars of this study are more metal-poor than the other aluminum enhanced stars. We test if NLTE effects might be responsible, but applying the NLTE corrections of \citet{Andrievsky08} increases the overabundance even more. It might be possible that (some of) the investigated RR\,Lyrae stars are underabundant in aluminum and therefore we were not able to detect the lines. Some of the LMC GC stars of \citet{Mucciarelli10} have estimates for aluminum. These stars, with metallicities of the BHB stars, show abundance patterns similar to the Galactic RR\,Lyrae stars of \citet{For11} and to the Galactic BHB stars. However, we cannot conclude that all RR\,Lyrae stars in the MCs are overabundant in [Al/Fe] and more stars need to be investigated.

\subsection{Iron-peak elements}

For the iron-peak elements we are only able to synthesize a few chromium lines for three LMC RR\,Lyrae stars to within acceptable error margins. In general these abundances are higher than in the comparison MW RR\,Lyrae stars. However, the low S/N of our spectra does not allow us to identify the lines properly and we recommend to take these values as upper limits.

\begin{figure}
\includegraphics[width=0.47\textwidth]{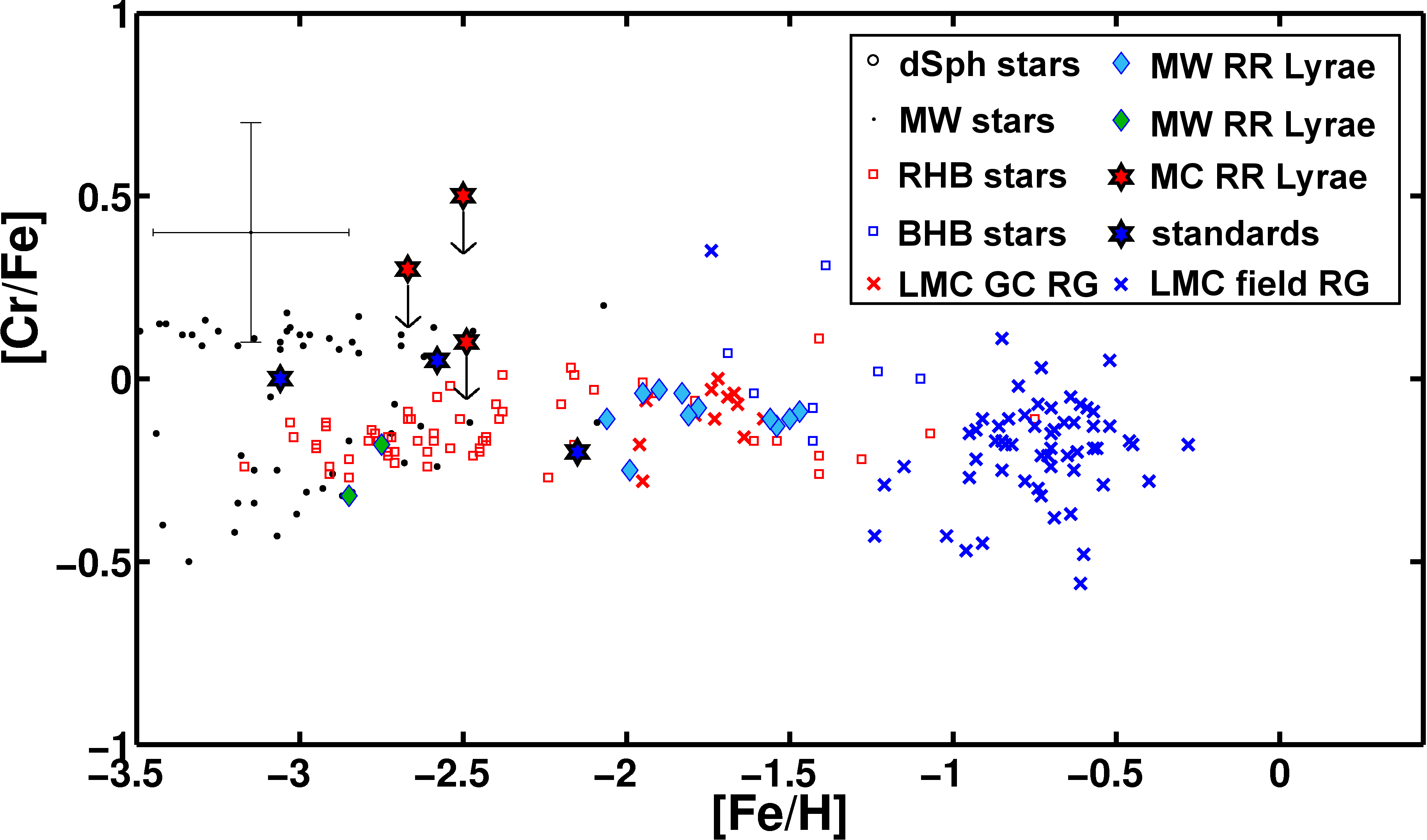} 
\caption{Chromium to iron ratio versus [Fe/H]. Only for three of our LMC RR\,Lyrae stars upper limits (indicated by the arrows) of chromium can be synthesized in our spectral synthesis analysis, but no other iron-peak elements. The abundances are in general a bit higher than the comparison values from MW RR\,Lyrae and HB stars. The color coding is the same as in Figure\,\ref{alpha_Fe}.}
\label{iron_peak}
\end{figure}

\subsection{Neutron-capture elements}

\begin{figure}
\includegraphics[width=0.47\textwidth]{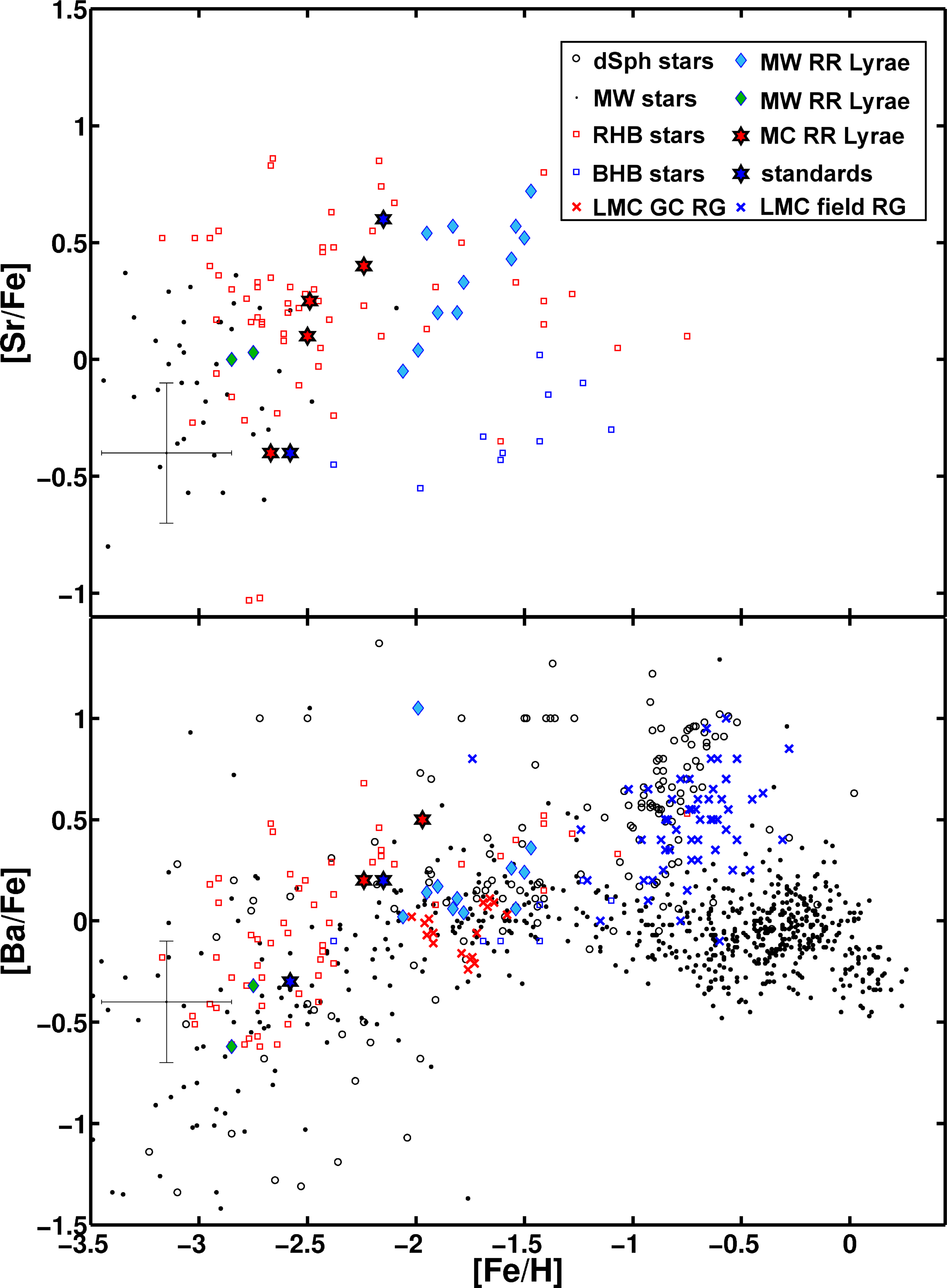} 
\caption{Strontium to iron and barium to iron ratios vs. [Fe/H]. There is good agreement with the literature data (see Figure\,\ref{alpha_Fe}). The strontium abundances of the LMC RR\,Lyrae stars fit well within the trend set by the MW RR\,Lyrae and HB stars (upper panel). For dSph and MW stars a significantly different trend of the barium to iron ratio is only observable at metallicities higher than those of our RR\,Lyrae sample (lower panel). Therefore we cannot distinguish whether the RR\,Lyrae stars follow the trend of the dSph or the MW stars.} 
\label{neutron_capture}
\end{figure}

Neutron-capture elements are represented by synthesized abundance measurements of strontium for four of our LMC RR\,Lyrae stars and synthesized abundances of barium for two of our LMC RR\,Lyrae stars. In Figure\,\ref{neutron_capture} the abundances are compared with the literature datasets mentioned in Section\,\ref{alpha}. 

For strontium a slight overabundance is found in most of the literature RR\,Lyrae and HB stars. However, there is no well defined trend and some stars are significantly underabundant. We find three RR\,Lyrae stars that are enhanced in strontium, while the abundance of one star deviates significantly from the other three. This is RR\,177, the same star that showed a low abundance pattern in the $\alpha$-elements. All stars are, however, in good agreement with the literature values, see Figure\,\ref{neutron_capture}. \citet{Mashonkina01} found that a NLTE correction of $\Delta_{\mathrm{NLTE}} = +0.6$ should be applied. Adding this value to our LTE measurements, the abundances of our RR\,Lyrae stars would still be in agreement with the abundances found for other Galactic RR\,Lyrae and RHB stars.

The barium content of our RR\,Lyrae stars is slightly overabundant suggestive of s-process contributions from AGB stars. Comparing it to the MW RR\,Lyrae stars by \citet{For11} and the RGs of the GCs \citep{Johnson06, Mucciarelli10}, we find that our abundances are a bit enhanced. However, within the uncertainties good agreement is found with these stars as well as with the RHB stars and the RGs of the dSphs (see Figure\,\ref{neutron_capture}). \citet{Andrievsky09} computed $\Delta_{\mathrm{NLTE}}$ corrections for the three lines at 4554\AA, 5853\AA \ and 6496\AA. Therefore two of the four lines used in this work have been investigated. For stars with temperatures above 6000\,K, as our RR\,Lyrae stars, a trend of higher barium abundances including NLTE corrections is found. This would enhance the abundance differences between the RGB stars of the GCs and the RR\,Lyrae stars. The hyperfine splitting \citep[HFS, e.g.,][]{McWilliam95, Short06} leads to overestimation of the barium content. Ignoring this effect could be an explanation for the slight overabundance of the RR\,Lyrae stars.

\subsection{The $\alpha$-poor star RR\,177}
\label{RR177}

For our most metal-poor star RR\,177, we find unusual abundances of the other elements. With $\mathrm{[Fe/H]} = -2.7$, this is the most metal-poor star currently known in the LMC. Therefore, we would expect the star to be similarly $\alpha$-enhanced as most of the metal-poor stars. Instead, we find $\mathrm{[Mg/Fe]} = -0.5$, with a signal well above the noise level. For [Ca/Fe] we do not find a conclusive value, but the spectral synthesis shows that the abundance has to be lower than $\mathrm{[Ca/Fe]} < 0$. For the $\alpha$-element titanium we find $\mathrm{[Ti/Fe]} = 0.0$, a bit higher than the abundances for the other $\alpha$-elements. However, as pointed out in \citet{Thielemann96} the production of titanium can also happen through complete silicon burning with an $\alpha$-rich freeze-out. The complete sequences inside a star that lead to the synthesis of titanium are not yet fully understood \citep[e.g.,][]{Woosley95}.

The low abundance of the $\alpha$-elements might be indicative for a single enrichment event without sufficient mixing of the supernova debris. According to \citet{Argast00, Argast02}, the mixing of the ejecta into a gas cloud of $10^4\,\mathrm{M}_{\odot}$ may lead to an $\alpha$-underabundance for individual stars. Following these models, the chromium abundance would appear rather normal, as observed for RR\,177 \citep[compare Figure 2 of][]{Argast00}. Conversely, a lower [$\alpha$/Fe] ratio may be caused by supernovae type\,Ia \textquoteleft pockets\textquoteright \ as described in the models of \citet{Marcolini08}.

To date, Galactic field and dSph stars with such an amount of selective depletion in the $\alpha$- (and possibly in some neutron-capture) elements have only been found at metallicities $\ga -2$\,dex \citep[e.g.,][]{Carney97, Ivans03, Koch08a}.

\section{Comparison of photometric and spectroscopic metallicities}
\label{comparison}

\begin{figure}
\includegraphics[width=0.47\textwidth]{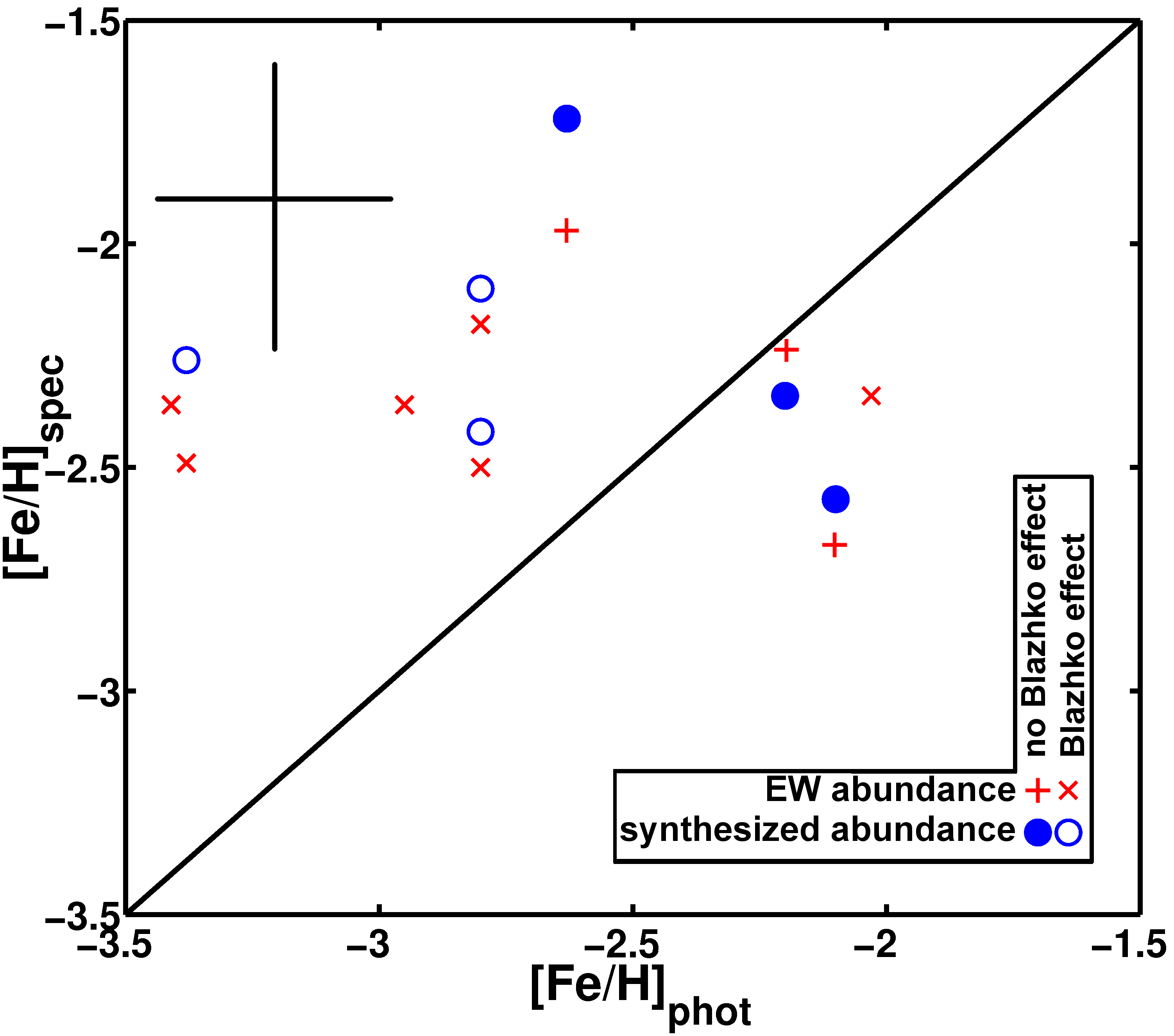} 
\caption{Photometric metallicity estimates versus spectroscopic measurements. The iron abundances from the EW measurements are labeled with the red crosses/pluses, while the synthesized abundances of iron are shown as blue circles. The filled circles and pluses are for stars with no detected Blazhko effect, while the open circles and crosses are the stars with a Blazhko effect. For the SMC RR\,Lyrae stars no synthesized value is available. Overall we see a trend for lower spectroscopic metallicities for stars with photometric metallicities around $-2$\,dex. For the RR\,Lyrae stars with [Fe/H]$_{\mathrm{phot}} < -2.5$\,dex the Blazhko effect might have altered the lightcurve and hence the Fourier parameters such that a lower metallicity is mimicked. Overall all target stars have similar spectroscopic metallicities. }
\label{Fe_phot_vs_spec}
\end{figure}

Comparing the photometric metallicity of the RR\,Lyrae stars obtained from the Fourier parameters of the lightcurves \citep{Haschke12_MDF} with the iron abundances of our spectra, we find trends as shown in Figure\,\ref{Fe_phot_vs_spec}. For the most metal-rich stars of our sample, i.e., the three stars with $\mathrm{[Fe/H]_{phot/J95}} \geq -2.20$\,dex, we find lower spectroscopic metallicities. These differences range from $-0.04$\,dex to $-0.57$\,dex. This discrepancy of too high photometric metallicity estimates for the metal-poor stars is a well known problem with the $V$ and $I$\,band calibration and was found to amount to 0.2\,dex to 0.3\,dex \citep[e.g.,][]{Dekany09, Arellano11} for Galactic RR\,Lyrae stars. The offset is attributed to the low number of metal-poor calibration stars for the Fourier decomposition method. \citet{Papadakis00} found that the metallicity scale of \citet{Zinn84} is about 0.3\,dex lower than that of \citet{Jurcsik95}. If we transform the photometric metallicities to the scale of \citet{Zinn84} and compare them to our spectroscopic metallicities we find very good agreement between both values with a mean difference of $0.02 \pm 0.14$\,dex.

However, for the RR\,Lyrae stars that may be even more metal-poor, we find a completely different picture. The stars with a photometric metallicity estimate of $\mathrm{[Fe/H]_{phot/J95}} \leq -2.80$\,dex turn out to be more metal-rich by $0.31$\,dex to $1.05$\,dex when analyzed spectroscopically. On the scale of \citet{Zinn84} the differences between the photometric and spectroscopic metallicities are even larger by 0.3\,dex. Comparing the spectroscopic metallicities of these two groups, we find that they are indistinguishable. On average both groups have a mean value of $\mathrm{[Fe/H]_{spec}} = -2.4$\,dex. 

\begin{figure}
\includegraphics[width=0.47\textwidth]{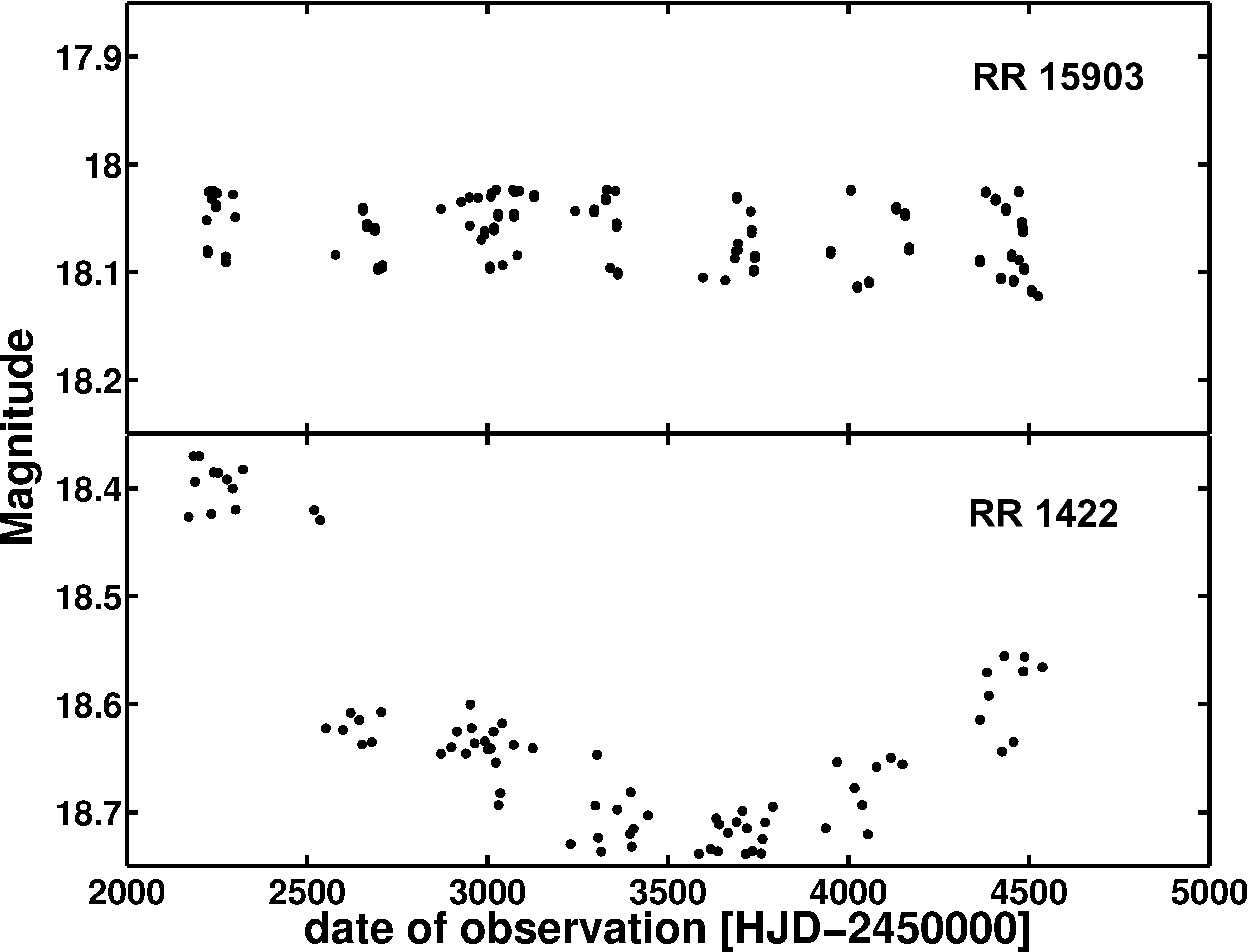} 
\caption{Variation of the magnitude at maximum light for the target stars RR\,15903 (upper panel) and RR\,1422 (lower panel). All observations of the OGLE team within a phase of 0.2 around the maximum light are plotted versus the date of observation. For RR\,15903 only scatter due to the magnitude variation caused by the phase width of $\Delta \Phi = 0.2$ is observed. The variation of the magnitude at presumable maximum light for RR\,1422 is an indication for the presence of an additional Blazhko effect.}
\label{lightcurve_modulation}
\end{figure}

The photometric metallicity estimate depends on the shape of the lightcurve, which may be influenced by the Blazhko effect \citep{Blazko07}. This additional modulation of the lightcurve, which changes the times ($t_{max}$) and the magnitude ($V_{max}$) of maximum brightness, happens on the order of $10-100$\,days and is believed to be present in up to $50\%$ of all RR\,Lyrae stars \citep[e.g.,][]{Jurcsik09}. In order to find the additional modulation of the lightcurve we analyze the OGLE\,III lightcurves of our target stars using the phase dispersion minimization (PDM) technique in the PDM2\footnote{www.stellingwerf.com} code \citep{Stellingwerf78}. The determined periods agree perfectly with the periods published by the OGLE team. The Blazhko amplitudes are much weaker than the fundamental pulsation and were therefore not detected by the PDM technique. Therefore we took all observations of OGLE, that were obtained within a phase interval of 0.2 around the maximum magnitude of an RR\,Lyrae target and plot them against their observation date (see Figure\,\ref{lightcurve_modulation}). For three RR\,Lyrae stars the observed magnitudes vary by less than 0.1\,mag within the time of observation (upper panel of Figure\,\ref{lightcurve_modulation}). If a periodic modulation is found, as shown in the lower panel of Figure\,\ref{lightcurve_modulation}, we assume the star to experience a Blazhko effect. This is the case for six of our nine target stars (Table\,\ref{list_RRL}).

Comparing the photometric and spectroscopic metallicity estimates of the RR\,Lyrae stars indicative of experiencing a Blazhko effect, a tight relation is found. The stars with extremely low photometric metallicities and therefore large discrepancies between their photometric and spectroscopic estimates all experience additional modulations of their lightcurves (open circles and crosses in Figure\,\ref{Fe_phot_vs_spec}). The stars with good agreement between their photometric and spectroscopic metallicities do not show evidence for the Blazhko effect (filled circles and pluses in Figure\,\ref{Fe_phot_vs_spec}). We conclude the lightcurves of the stars with very low photometric metallicities are influenced by the Blazhko effect. This additional modulation appears to alter the Fourier parameters such that an extremely metal-poor star is mimicked. This affects the five most metal-poor candidates of our target stars (in terms of photometric metallicities) in our sample. We suspect that in addition other effects may play a role, such as the limited metallicity range of the calibrating stars as already suggested by other authors. Nonetheless, the Blazhko effect would seem to be a major culprit for the discrepancy between the photometric and the spectroscopic metallicities.

Comparing the deviations of the temperature estimates, we do not find a trend that stars with an indication of a Blazhko effect have larger differences in their temperatures. We conclude that the temperature estimation is in our case more sensitive to the sampling of the lightcurve than to the Blazhko effect.

\section{Conclusions}
\label{Conclusions}

We present the first detailed spectroscopic analysis of Magellanic Cloud RR\,Lyrae stars, six of which are located in the LMC and three stars in the SMC. With metallicities as low as $\mathrm{[Fe/H]} = -2.7$\,dex, we find the most metal-poor stars yet known in the MCs. A mean metallicity of $\mathrm{[Fe/H]} = -2.4$\,dex for our nine objects allows us to probe the earliest currently available tracers of the chemical evolution of the MCs.

The candidates for the most metal-poor RR\,Lyrae stars were identified via photometric metallicity estimates \citep{Haschke12_MDF}. These candidates generally turn out to be more metal-rich when using spectroscopic measures. Overall, the metallicities of our RR\,Lyrae stars are significantly higher than those of the extremely metal-poor RGB stars found in several dSphs \citep[e.g.,][]{Frebel10a, Norris10a} or the most metal-poor stars of the Galactic halo \citep[e.g.,][]{Frebel05, Caffau11}. The reason for this lack of extremely metal-poor RR\,Lyrae stars are the evolutionary tracks of stars beyond the main sequence. As pointed out by \citet{Yoon02} extremely metal-poor stars do not enter the instability strip and therefore one would not expect to find RR\,Lyrae stars with $\mathrm{[Fe/H]} \leq -3$\,dex.

For the $\alpha$-enhancement an overabundance between $0.1 < [\alpha/\mathrm{Fe}] < 0.5$\,dex is found for Galactic RR\,Lyrae stars \citep{For11, Hansen11} as well as for red giants in the GCs of the MCs \citep{Johnson06, Mucciarelli09, Mucciarelli10}, in dSphs \citep[e.g.,][and references mentioned above]{Koch08a} and in the MW halo \citep[e.g.,][and references therein]{Venn04}. Our spectral synthesis of the $\alpha$-elements in our metal-poor field RR\,Lyrae stars in the MCs reveals a mean $\alpha$-enhancement of $0.36 \pm 0.25$\,dex in very good agreement with these other sources. 

For the odd-Z elements we find a small overabundance in sodium and a significant overabundance in aluminum in good agreement with the red giants of the MC GCs and the Galactic RR\,Lyrae stars. The upper limits for our measured iron-peak elements are consistent with the abundances determined for different types of Galactic stars of similar metallicity \citep[e.g.,][]{Preston06, For10}. For the neutron capture elements we also find good agreement between our abundances and the values obtained for the different dSphs, the LMC GCs and the halo stars of the MW. 

As members of the old population of the MC our metal-poor RR\,Lyrae stars can be expected to have been primarily enriched by supernovae of type\,II, with little contribution from supernovae of type\,Ia. The abundance signatures largely confirm this assumption. Our results are, therefore, consistent with the model of an invariant, massive star, initial mass function (IMF) for all environments at early times \citep[e.g.,][]{Wyse02, Frebel10a}.

The spread of abundances observed in our target stars and the $\alpha$-element deficient star RR\,177 are indicative of local star formation in bursts in an inhomogeneously enriched environment or even enrichment by single supernova Ia events. Therefore we expect that star formation happened locally in a stochastical way and that the early MCs were not chemically well mixed. This is also consistent with the range of metallicities found in SMC star clusters at any given age \citep[e.g.,][]{Glatt08b} as well as with findings for (old) field populations in dSphs \citep[e.g.,][]{Koch08b}. 

In general we find that the abundances of the old and metal-poor MCs RR\,Lyrae stars are consistent with the abundances measured in the MW halo. Recent numerical simulations \citep{Font06,  Robertson05} indicate that the halo was primarily built up from early mergers of the MW with MC-sized objects. From our results we can confirm that the abundance patterns of all investigated elements of the metal-poor, old MC stars and the MW halo are compatible. Therefore it is conceivable that early mergers of \textquotedblleft proto-MCs\textquotedblright \ with the MW contributed to the build-up of its stellar halo.


\acknowledgements  
R.~H. is grateful to the Heidelberg Graduate School for Fundamental Physics (HGSFP)(grant number 129/1) for providing support for a scientific visit at Harvard University.
E.K.~G, S.~D. \& C.J.~H. acknowledge support by the Sonderforschungbereich SFB881 (``The Milky Way System'') of the Deutsche Forschungsgemeinschaft (DFG) in the subprojects A2, A3 and A5.
A.~F. is supported by a Clay Fellowship administered by the Smithsonian Astrophysical Observatory.
A.~K. acknowledges the DFG for funding from Emmy-Noether grant Ko 4161/1.

{\it Facilities:}\facility{Magellan:MAGE}

\bibliography{Bibliography.bib}

\begin{thebibliography}{113}
\expandafter\ifx\csname natexlab\endcsname\relax\def\natexlab#1{#1}\fi

\bibitem[{{Ad{\'e}n} {et~al.}(2011){Ad{\'e}n}, {Eriksson}, {Feltzing},
  {Grebel}, {Koch}, \& {Wilkinson}}]{Aden11}
{Ad{\'e}n}, D., {Eriksson}, K., {Feltzing}, S., {Grebel}, E.~K., {Koch}, A., \&
  {Wilkinson}, M.~I. 2011, \aap, 525, A153

\bibitem[{{Anders} \& {Grevesse}(1989)}]{Anders89}
{Anders}, E. \& {Grevesse}, N. 1989, \gca, 53, 197

\bibitem[{{Andrievsky} {et~al.}(2008){Andrievsky}, {Spite}, {Korotin}, {Spite},
  {Bonifacio}, {Cayrel}, {Hill}, \& {Fran{\c c}ois}}]{Andrievsky08}
{Andrievsky}, S.~M., {Spite}, M., {Korotin}, S.~A., {Spite}, F., {Bonifacio},
  P., {Cayrel}, R., {Hill}, V., \& {Fran{\c c}ois}, P. 2008, \aap, 481, 481

\bibitem[{{Andrievsky} {et~al.}(2009){Andrievsky}, {Spite}, {Korotin}, {Spite},
  {Fran{\c c}ois}, {Bonifacio}, {Cayrel}, \& {Hill}}]{Andrievsky09}
{Andrievsky}, S.~M., {Spite}, M., {Korotin}, S.~A., {Spite}, F., {Fran{\c
  c}ois}, P., {Bonifacio}, P., {Cayrel}, R., \& {Hill}, V. 2009, \aap, 494,
  1083

\bibitem[{{Aoki} {et~al.}(2009){Aoki}, {Arimoto}, {Sadakane}, {Tolstoy},
  {Battaglia}, {Jablonka}, {Shetrone}, {Letarte}, {Irwin}, {Hill}, {Francois},
  {Venn}, {Primas}, {Helmi}, {Kaufer}, {Tafelmeyer}, {Szeifert}, \&
  {Babusiaux}}]{Aoki09}
{Aoki}, W., {Arimoto}, N., {Sadakane}, K., {Tolstoy}, E., {Battaglia}, G.,
  {Jablonka}, P., {Shetrone}, M., {Letarte}, B., {Irwin}, M., {Hill}, V.,
  {Francois}, P., {Venn}, K., {Primas}, F., {Helmi}, A., {Kaufer}, A.,
  {Tafelmeyer}, M., {Szeifert}, T., \& {Babusiaux}, C. 2009, \aap, 502, 569

\bibitem[{{Arellano Ferro} {et~al.}(2011){Arellano Ferro}, {Figuera Jaimes},
  {Giridhar}, {Bramich}, {Hern{\'a}ndez Santisteban}, \&
  {Kuppuswamy}}]{Arellano11}
{Arellano Ferro}, A., {Figuera Jaimes}, R., {Giridhar}, S., {Bramich}, D.~M.,
  {Hern{\'a}ndez Santisteban}, J.~V., \& {Kuppuswamy}, K. 2011, \mnras, 416,
  2265

\bibitem[{{Argast} {et~al.}(2000){Argast}, {Samland}, {Gerhard}, \&
  {Thielemann}}]{Argast00}
{Argast}, D., {Samland}, M., {Gerhard}, O.~E., \& {Thielemann}, F.-K. 2000,
  \aap, 356, 873

\bibitem[{{Argast} {et~al.}(2002){Argast}, {Samland}, {Thielemann}, \&
  {Gerhard}}]{Argast02}
{Argast}, D., {Samland}, M., {Thielemann}, F.-K., \& {Gerhard}, O.~E. 2002,
  \aap, 388, 842

\bibitem[{{Beveridge} \& {Sneden}(1994)}]{Beveridge94}
{Beveridge}, R.~C. \& {Sneden}, C. 1994, \aj, 108, 285

\bibitem[{{Bla{\v z}ko}(1907)}]{Blazko07}
{Bla{\v z}ko}, S. 1907, Astronomische Nachrichten, 175, 325

\bibitem[{{Boesgaard} \& {Novicki}(2006)}]{Boesgaard06}
{Boesgaard}, A.~M. \& {Novicki}, M.~C. 2006, \apj, 641, 1122

\bibitem[{{Bonifacio} {et~al.}(2009){Bonifacio}, {Spite}, {Cayrel}, {Hill},
  {Spite}, {Fran{\c c}ois}, {Plez}, {Ludwig}, {Caffau}, {Molaro}, {Depagne},
  {Andersen}, {Barbuy}, {Beers}, {Nordstr{\"o}m}, \& {Primas}}]{Bonifacio09}
{Bonifacio}, P., {Spite}, M., {Cayrel}, R., {Hill}, V., {Spite}, F., {Fran{\c
  c}ois}, P., {Plez}, B., {Ludwig}, H.-G., {Caffau}, E., {Molaro}, P.,
  {Depagne}, E., {Andersen}, J., {Barbuy}, B., {Beers}, T.~C., {Nordstr{\"o}m},
  B., \& {Primas}, F. 2009, \aap, 501, 519

\bibitem[{{Borissova} {et~al.}(2006){Borissova}, {Minniti}, {Rejkuba}, \&
  {Alves}}]{Borissova06}
{Borissova}, J., {Minniti}, D., {Rejkuba}, M., \& {Alves}, D. 2006, \aap, 460,
  459

\bibitem[{{Borissova} {et~al.}(2004){Borissova}, {Minniti}, {Rejkuba}, {Alves},
  {Cook}, \& {Freeman}}]{Borissova04}
{Borissova}, J., {Minniti}, D., {Rejkuba}, M., {Alves}, D., {Cook}, K.~H., \&
  {Freeman}, K.~C. 2004, \aap, 423, 97

\bibitem[{{Butler} {et~al.}(1982){Butler}, {Demarque}, \& {Smith}}]{Butler82}
{Butler}, D., {Demarque}, P., \& {Smith}, H.~A. 1982, \apj, 257, 592

\bibitem[{{Caffau} {et~al.}(2011){Caffau}, {Bonifacio}, {Fran{\c c}ois},
  {Sbordone}, {Monaco}, {Spite}, {Spite}, {Ludwig}, {Cayrel}, {Zaggia},
  {Hammer}, {Randich}, {Molaro}, \& {Hill}}]{Caffau11}
{Caffau}, E., {Bonifacio}, P., {Fran{\c c}ois}, P., {Sbordone}, L., {Monaco},
  L., {Spite}, M., {Spite}, F., {Ludwig}, H.-G., {Cayrel}, R., {Zaggia}, S.,
  {Hammer}, F., {Randich}, S., {Molaro}, P., \& {Hill}, V. 2011, \nat, 477, 67

\bibitem[{{Carney} {et~al.}(1997){Carney}, {Wright}, {Sneden}, {Laird},
  {Aguilar}, \& {Latham}}]{Carney97}
{Carney}, B.~W., {Wright}, J.~S., {Sneden}, C., {Laird}, J.~B., {Aguilar},
  L.~A., \& {Latham}, D.~W. 1997, \aj, 114, 363

\bibitem[{{Carrera} {et~al.}(2008{\natexlab{a}}){Carrera}, {Gallart},
  {Aparicio}, {Costa}, {M{\'e}ndez}, \& {No{\"e}l}}]{Carrera08b}
{Carrera}, R., {Gallart}, C., {Aparicio}, A., {Costa}, E., {M{\'e}ndez}, R.~A.,
  \& {No{\"e}l}, N.~E.~D. 2008{\natexlab{a}}, \aj, 136, 1039

\bibitem[{{Carrera} {et~al.}(2011){Carrera}, {Gallart}, {Aparicio}, \&
  {Hardy}}]{Carrera11}
{Carrera}, R., {Gallart}, C., {Aparicio}, A., \& {Hardy}, E. 2011, \aj, 142, 61

\bibitem[{{Carrera} {et~al.}(2008{\natexlab{b}}){Carrera}, {Gallart}, {Hardy},
  {Aparicio}, \& {Zinn}}]{Carrera08a}
{Carrera}, R., {Gallart}, C., {Hardy}, E., {Aparicio}, A., \& {Zinn}, R.
  2008{\natexlab{b}}, \aj, 135, 836

\bibitem[{{Casagrande} {et~al.}(2010){Casagrande}, {Ram{\'{\i}}rez},
  {Mel{\'e}ndez}, {Bessell}, \& {Asplund}}]{Casagrande10}
{Casagrande}, L., {Ram{\'{\i}}rez}, I., {Mel{\'e}ndez}, J., {Bessell}, M., \&
  {Asplund}, M. 2010, \aap, 512, 54

\bibitem[{{Castelli} {et~al.}(1997){Castelli}, {Gratton}, \&
  {Kurucz}}]{Castelli97}
{Castelli}, F., {Gratton}, R.~G., \& {Kurucz}, R.~L. 1997, \aap, 318, 841

\bibitem[{{Cayrel} {et~al.}(2004){Cayrel}, {Depagne}, {Spite}, {Hill}, {Spite},
  {Fran{\c c}ois}, {Plez}, {Beers}, {Primas}, {Andersen}, {Barbuy},
  {Bonifacio}, {Molaro}, \& {Nordstr{\"o}m}}]{Cayrel04}
{Cayrel}, R., {Depagne}, E., {Spite}, M., {Hill}, V., {Spite}, F., {Fran{\c
  c}ois}, P., {Plez}, B., {Beers}, T., {Primas}, F., {Andersen}, J., {Barbuy},
  B., {Bonifacio}, P., {Molaro}, P., \& {Nordstr{\"o}m}, B. 2004, \aap, 416,
  1117

\bibitem[{{Cenarro} {et~al.}(2007){Cenarro}, {Peletier},
  {S{\'a}nchez-Bl{\'a}zquez}, {Selam}, {Toloba}, {Cardiel},
  {Falc{\'o}n-Barroso}, {Gorgas}, {Jim{\'e}nez-Vicente}, \&
  {Vazdekis}}]{Cenarro07}
{Cenarro}, A.~J., {Peletier}, R.~F., {S{\'a}nchez-Bl{\'a}zquez}, P., {Selam},
  S.~O., {Toloba}, E., {Cardiel}, N., {Falc{\'o}n-Barroso}, J., {Gorgas}, J.,
  {Jim{\'e}nez-Vicente}, J., \& {Vazdekis}, A. 2007, \mnras, 374, 664

\bibitem[{{Clementini} {et~al.}(1995){Clementini}, {Carretta}, {Gratton},
  {Merighi}, {Mould}, \& {McCarthy}}]{Clementini95}
{Clementini}, G., {Carretta}, E., {Gratton}, R., {Merighi}, R., {Mould}, J.~R.,
  \& {McCarthy}, J.~K. 1995, \aj, 110, 2319

\bibitem[{{Cohen} \& {Huang}(2009)}]{Cohen09}
{Cohen}, J.~G. \& {Huang}, W. 2009, \apj, 701, 1053

\bibitem[{{Cohen} \& {Huang}(2010)}]{Cohen10}
---. 2010, \apj, 719, 931

\bibitem[{{Cole} {et~al.}(2005){Cole}, {Tolstoy}, {Gallagher}, \&
  {Smecker-Hane}}]{Cole05}
{Cole}, A.~A., {Tolstoy}, E., {Gallagher}, III, J.~S., \& {Smecker-Hane}, T.~A.
  2005, \aj, 129, 1465

\bibitem[{{Da Costa} \& {Hatzidimitriou}(1998)}]{DaCosta98}
{Da Costa}, G.~S. \& {Hatzidimitriou}, D. 1998, \aj, 115, 1934

\bibitem[{{De Propris} {et~al.}(2010){De Propris}, {Rich}, {Mallery}, \&
  {Howard}}]{DePropris10}
{De Propris}, R., {Rich}, R.~M., {Mallery}, R.~C., \& {Howard}, C.~D. 2010,
  \apjl, 714, L249

\bibitem[{{de Vaucouleurs} \& {Freeman}(1972)}]{Vaucouleurs72}
{de Vaucouleurs}, G. \& {Freeman}, K.~C. 1972, Vistas in Astronomy, 14, 163

\bibitem[{{D{\'e}k{\'a}ny} \& {Kov{\'a}cs}(2009)}]{Dekany09}
{D{\'e}k{\'a}ny}, I. \& {Kov{\'a}cs}, G. 2009, \aap, 507, 803

\bibitem[{{Feltzing} {et~al.}(2009){Feltzing}, {Eriksson}, {Kleyna}, \&
  {Wilkinson}}]{Feltzing09}
{Feltzing}, S., {Eriksson}, K., {Kleyna}, J., \& {Wilkinson}, M.~I. 2009, \aap,
  508, L1

\bibitem[{{Fernley} \& {Barnes}(1996)}]{Fernley96}
{Fernley}, J. \& {Barnes}, T.~G. 1996, \aap, 312, 957

\bibitem[{{Font} {et~al.}(2006){Font}, {Johnston}, {Bullock}, \&
  {Robertson}}]{Font06}
{Font}, A.~S., {Johnston}, K.~V., {Bullock}, J.~S., \& {Robertson}, B.~E. 2006,
  \apj, 638, 585

\bibitem[{{For} \& {Sneden}(2010)}]{For10}
{For}, B.-Q. \& {Sneden}, C. 2010, \aj, 140, 1694

\bibitem[{{For} {et~al.}(2011){For}, {Sneden}, \& {Preston}}]{For11}
{For}, B.-Q., {Sneden}, C., \& {Preston}, G.~W. 2011, \apjs, 197, 29

\bibitem[{{Frebel} {et~al.}(2005){Frebel}, {Aoki}, {Christlieb}, {Ando},
  {Asplund}, {Barklem}, {Beers}, {Eriksson}, {Fechner}, {Fujimoto}, {Honda},
  {Kajino}, {Minezaki}, {Nomoto}, {Norris}, {Ryan}, {Takada-Hidai},
  {Tsangarides}, \& {Yoshii}}]{Frebel05}
{Frebel}, A., {Aoki}, W., {Christlieb}, N., {Ando}, H., {Asplund}, M.,
  {Barklem}, P.~S., {Beers}, T.~C., {Eriksson}, K., {Fechner}, C., {Fujimoto},
  M.~Y., {Honda}, S., {Kajino}, T., {Minezaki}, T., {Nomoto}, K., {Norris},
  J.~E., {Ryan}, S.~G., {Takada-Hidai}, M., {Tsangarides}, S., \& {Yoshii}, Y.
  2005, \nat, 434, 871

\bibitem[{{Frebel} {et~al.}(2010{\natexlab{a}}){Frebel}, {Kirby}, \&
  {Simon}}]{Frebel10a}
{Frebel}, A., {Kirby}, E.~N., \& {Simon}, J.~D. 2010{\natexlab{a}}, \nat, 464,
  72

\bibitem[{{Frebel} {et~al.}(2010{\natexlab{b}}){Frebel}, {Simon}, {Geha}, \&
  {Willman}}]{Frebel10b}
{Frebel}, A., {Simon}, J.~D., {Geha}, M., \& {Willman}, B. 2010{\natexlab{b}},
  \apj, 708, 560

\bibitem[{{Fulbright}(2000)}]{Fulbright00}
{Fulbright}, J.~P. 2000, \aj, 120, 1841

\bibitem[{{Glatt} {et~al.}(2008{\natexlab{a}}){Glatt}, {Gallagher}, {Grebel},
  {Nota}, {Sabbi}, {Sirianni}, {Clementini}, {Tosi}, {Harbeck}, {Koch}, \&
  {Cracraft}}]{Glatt08a}
{Glatt}, K., {Gallagher}, III, J.~S., {Grebel}, E.~K., {Nota}, A., {Sabbi}, E.,
  {Sirianni}, M., {Clementini}, G., {Tosi}, M., {Harbeck}, D., {Koch}, A., \&
  {Cracraft}, M. 2008{\natexlab{a}}, \aj, 135, 1106

\bibitem[{{Glatt} {et~al.}(2008{\natexlab{b}}){Glatt}, {Grebel}, {Sabbi},
  {Gallagher}, {Nota}, {Sirianni}, {Clementini}, {Tosi}, {Harbeck}, {Koch},
  {Kayser}, \& {Da Costa}}]{Glatt08b}
{Glatt}, K., {Grebel}, E.~K., {Sabbi}, E., {Gallagher}, J.~S., {Nota}, A.,
  {Sirianni}, M., {Clementini}, G., {Tosi}, M., {Harbeck}, D., {Koch}, A.,
  {Kayser}, A., \& {Da Costa}, G. 2008{\natexlab{b}}, \aj, 136, 1703

\bibitem[{{Gonidakis} {et~al.}(2009){Gonidakis}, {Livanou}, {Kontizas},
  {Klein}, {Kontizas}, {Belcheva}, {Tsalmantza}, \& {Karampelas}}]{Gonidakis09}
{Gonidakis}, I., {Livanou}, E., {Kontizas}, E., {Klein}, U., {Kontizas}, M.,
  {Belcheva}, M., {Tsalmantza}, P., \& {Karampelas}, A. 2009, \aap, 496, 375

\bibitem[{{Gratton} {et~al.}(2004){Gratton}, {Bragaglia}, {Clementini},
  {Carretta}, {Di Fabrizio}, {Maio}, \& {Taribello}}]{Gratton04}
{Gratton}, R.~G., {Bragaglia}, A., {Clementini}, G., {Carretta}, E., {Di
  Fabrizio}, L., {Maio}, M., \& {Taribello}, E. 2004, \aap, 421, 937

\bibitem[{{Grebel} \& {Gallagher}(2004)}]{Grebel04}
{Grebel}, E.~K. \& {Gallagher}, III, J.~S. 2004, \apjl, 610, 89

\bibitem[{{Grevesse} \& {Sauval}(1998)}]{Grevesse98}
{Grevesse}, N. \& {Sauval}, A.~J. 1998, \ssr, 85, 161

\bibitem[{{Hansen} {et~al.}(2011){Hansen}, {Nordstr{\"o}m}, {Bonifacio},
  {Spite}, {Andersen}, {Beers}, {Cayrel}, {Spite}, {Molaro}, {Barbuy},
  {Depagne}, {Fran{\c c}ois}, {Hill}, {Plez}, \& {Sivarani}}]{Hansen11}
{Hansen}, C.~J., {Nordstr{\"o}m}, B., {Bonifacio}, P., {Spite}, M., {Andersen},
  J., {Beers}, T.~C., {Cayrel}, R., {Spite}, F., {Molaro}, P., {Barbuy}, B.,
  {Depagne}, E., {Fran{\c c}ois}, P., {Hill}, V., {Plez}, B., \& {Sivarani}, T.
  2011, \aap, 527, A65

\bibitem[{{Harris} \& {Zaritsky}(2006)}]{Harris06}
{Harris}, J. \& {Zaritsky}, D. 2006, \aj, 131, 2514

\bibitem[{{Haschke} {et~al.}(2012){Haschke}, {Grebel}, {Duffau}, \&
  {Jin}}]{Haschke12_MDF}
{Haschke}, R., {Grebel}, E.~K., {Duffau}, S., \& {Jin}, S. 2012, \aj, 143, 48

\bibitem[{{Ivans} {et~al.}(2003){Ivans}, {Sneden}, {James}, {Preston},
  {Fulbright}, {H{\"o}flich}, {Carney}, \& {Wheeler}}]{Ivans03}
{Ivans}, I.~I., {Sneden}, C., {James}, C.~R., {Preston}, G.~W., {Fulbright},
  J.~P., {H{\"o}flich}, P.~A., {Carney}, B.~W., \& {Wheeler}, J.~C. 2003, \apj,
  592, 906

\bibitem[{{Johnson} {et~al.}(2006){Johnson}, {Ivans}, \& {Stetson}}]{Johnson06}
{Johnson}, J.~A., {Ivans}, I.~I., \& {Stetson}, P.~B. 2006, \apj, 640, 801

\bibitem[{{Jurcsik}(1995)}]{Jurcsik95}
{Jurcsik}, J. 1995, Acta Astronomica, 45, 653

\bibitem[{{Jurcsik} {et~al.}(2009){Jurcsik}, {S{\'o}dor}, {Szeidl}, {Hurta},
  {V{\'a}radi}, {Posztob{\'a}nyi}, {Vida}, {Hajdu}, {K{\H o}v{\'a}ri}, {Nagy},
  {Moln{\'a}r}, \& {Belucz}}]{Jurcsik09}
{Jurcsik}, J., {S{\'o}dor}, {\'A}., {Szeidl}, B., {Hurta}, Z., {V{\'a}radi},
  M., {Posztob{\'a}nyi}, K., {Vida}, K., {Hajdu}, G., {K{\H o}v{\'a}ri}, Z.,
  {Nagy}, I., {Moln{\'a}r}, L., \& {Belucz}, B. 2009, \mnras, 400, 1006

\bibitem[{{Kayser} {et~al.}(2007){Kayser}, {Grebel}, {Harbeck}, {Cole}, {Koch},
  {Glatt}, {Gallagher}, \& {da Costa}}]{Kayser07}
{Kayser}, A., {Grebel}, E.~K., {Harbeck}, D.~R., {Cole}, A.~A., {Koch}, A.,
  {Glatt}, K., {Gallagher}, J.~S., \& {da Costa}, G.~S. 2007, in IAU Symposium
  241, ed. {A.~Vazdekis \& R.~F.~Peletier} (Cambridge University Press,
  Cambridge), 351

\bibitem[{{Koch} {et~al.}(2008{\natexlab{a}}){Koch}, {Grebel}, {Gilmore},
  {Wyse}, {Kleyna}, {Harbeck}, {Wilkinson}, \& {Wyn Evans}}]{Koch08a}
{Koch}, A., {Grebel}, E.~K., {Gilmore}, G.~F., {Wyse}, R.~F.~G., {Kleyna},
  J.~T., {Harbeck}, D.~R., {Wilkinson}, M.~I., \& {Wyn Evans}, N.
  2008{\natexlab{a}}, \aj, 135, 1580

\bibitem[{{Koch} {et~al.}(2008{\natexlab{b}}){Koch}, {McWilliam}, {Grebel},
  {Zucker}, \& {Belokurov}}]{Koch08b}
{Koch}, A., {McWilliam}, A., {Grebel}, E.~K., {Zucker}, D.~B., \& {Belokurov},
  V. 2008{\natexlab{b}}, \apjl, 688, L13

\bibitem[{{Kov{\'a}cs} \& {Zsoldos}(1995)}]{Kovacs95}
{Kov{\'a}cs}, G. \& {Zsoldos}, E. 1995, \aap, 293, L57

\bibitem[{{Lai} {et~al.}(2008){Lai}, {Bolte}, {Johnson}, {Lucatello}, {Heger},
  \& {Woosley}}]{Lai08}
{Lai}, D.~K., {Bolte}, M., {Johnson}, J.~A., {Lucatello}, S., {Heger}, A., \&
  {Woosley}, S.~E. 2008, \apj, 681, 1524

\bibitem[{{Lambert} {et~al.}(1996){Lambert}, {Heath}, {Lemke}, \&
  {Drake}}]{Lambert96}
{Lambert}, D.~L., {Heath}, J.~E., {Lemke}, M., \& {Drake}, J. 1996, \apjs, 103,
  183

\bibitem[{{Layden}(1994)}]{Layden94}
{Layden}, A.~C. 1994, \aj, 108, 1016

\bibitem[{{Layden}(1998)}]{Layden98}
---. 1998, \aj, 115, 193

\bibitem[{{Letarte} {et~al.}(2010){Letarte}, {Hill}, {Tolstoy}, {Jablonka},
  {Shetrone}, {Venn}, {Spite}, {Irwin}, {Battaglia}, {Helmi}, {Primas},
  {Fran{\c c}ois}, {Kaufer}, {Szeifert}, {Arimoto}, \& {Sadakane}}]{Letarte10}
{Letarte}, B., {Hill}, V., {Tolstoy}, E., {Jablonka}, P., {Shetrone}, M.,
  {Venn}, K.~A., {Spite}, M., {Irwin}, M.~J., {Battaglia}, G., {Helmi}, A.,
  {Primas}, F., {Fran{\c c}ois}, P., {Kaufer}, A., {Szeifert}, T., {Arimoto},
  N., \& {Sadakane}, K. 2010, \aap, 523, A17

\bibitem[{{Marcolini} {et~al.}(2008){Marcolini}, {D'Ercole}, {Battaglia}, \&
  {Gibson}}]{Marcolini08}
{Marcolini}, A., {D'Ercole}, A., {Battaglia}, G., \& {Gibson}, B.~K. 2008,
  \mnras, 386, 2173

\bibitem[{{Marshall} {et~al.}(2008){Marshall}, {Burles}, {Thompson},
  {Shectman}, {Bigelow}, {Burley}, {Birk}, {Estrada}, {Jones}, {Smith},
  {Kowal}, {Castillo}, {Storts}, \& {Ortiz}}]{Marshall08}
{Marshall}, J.~L., {Burles}, S., {Thompson}, I.~B., {Shectman}, S.~A.,
  {Bigelow}, B.~C., {Burley}, G., {Birk}, C., {Estrada}, J., {Jones}, P.,
  {Smith}, M., {Kowal}, V., {Castillo}, J., {Storts}, R., \& {Ortiz}, G. 2008,
  in Society of Photo-Optical Instrumentation Engineers (SPIE) Conference
  Series, ed. {Ian S. McLean \& Mark M. Casali}, Vol. 7014, 169

\bibitem[{{Mashonkina} \& {Gehren}(2001)}]{Mashonkina01}
{Mashonkina}, L. \& {Gehren}, T. 2001, \aap, 376, 232

\bibitem[{{McWilliam} {et~al.}(1995){McWilliam}, {Preston}, {Sneden}, \&
  {Searle}}]{McWilliam95}
{McWilliam}, A., {Preston}, G.~W., {Sneden}, C., \& {Searle}, L. 1995, \aj,
  109, 2757

\bibitem[{{Minniti} {et~al.}(2003){Minniti}, {Borissova}, {Rejkuba}, {Alves},
  {Cook}, \& {Freeman}}]{Minniti03}
{Minniti}, D., {Borissova}, J., {Rejkuba}, M., {Alves}, D.~R., {Cook}, K.~H.,
  \& {Freeman}, K.~C. 2003, Science, 301, 1508

\bibitem[{{Monaco} {et~al.}(2005){Monaco}, {Bellazzini}, {Bonifacio},
  {Ferraro}, {Marconi}, {Pancino}, {Sbordone}, \& {Zaggia}}]{Monaco05}
{Monaco}, L., {Bellazzini}, M., {Bonifacio}, P., {Ferraro}, F.~R., {Marconi},
  G., {Pancino}, E., {Sbordone}, L., \& {Zaggia}, S. 2005, \aap, 441, 141

\bibitem[{{Mucciarelli} {et~al.}(2010){Mucciarelli}, {Origlia}, \&
  {Ferraro}}]{Mucciarelli10}
{Mucciarelli}, A., {Origlia}, L., \& {Ferraro}, F.~R. 2010, \apj, 717, 277

\bibitem[{{Mucciarelli} {et~al.}(2009){Mucciarelli}, {Origlia}, {Ferraro}, \&
  {Pancino}}]{Mucciarelli09}
{Mucciarelli}, A., {Origlia}, L., {Ferraro}, F.~R., \& {Pancino}, E. 2009,
  \apjl, 695, L134

\bibitem[{{Norris} {et~al.}(2010{\natexlab{a}}){Norris}, {Gilmore}, {Wyse},
  {Yong}, \& {Frebel}}]{Norris10b}
{Norris}, J.~E., {Gilmore}, G., {Wyse}, R.~F.~G., {Yong}, D., \& {Frebel}, A.
  2010{\natexlab{a}}, \apjl, 722, L104

\bibitem[{{Norris} {et~al.}(2010{\natexlab{b}}){Norris}, {Wyse}, {Gilmore},
  {Yong}, {Frebel}, {Wilkinson}, {Belokurov}, \& {Zucker}}]{Norris10c}
{Norris}, J.~E., {Wyse}, R.~F.~G., {Gilmore}, G., {Yong}, D., {Frebel}, A.,
  {Wilkinson}, M.~I., {Belokurov}, V., \& {Zucker}, D.~B. 2010{\natexlab{b}},
  \apj, 723, 1632

\bibitem[{{Norris} {et~al.}(2010{\natexlab{c}}){Norris}, {Yong}, {Gilmore}, \&
  {Wyse}}]{Norris10a}
{Norris}, J.~E., {Yong}, D., {Gilmore}, G., \& {Wyse}, R.~F.~G.
  2010{\natexlab{c}}, \apj, 711, 350

\bibitem[{{Oke}(1966)}]{Oke66}
{Oke}, J.~B. 1966, \apj, 145, 468

\bibitem[{{Olsen} \& {Massey}(2007)}]{Olsen07}
{Olsen}, K.~A.~G. \& {Massey}, P. 2007, \apjl, 656, L61

\bibitem[{{Olsen} {et~al.}(2011){Olsen}, {Zaritsky}, {Blum}, {Boyer}, \&
  {Gordon}}]{Olsen11}
{Olsen}, K.~A.~G., {Zaritsky}, D., {Blum}, R.~D., {Boyer}, M.~L., \& {Gordon},
  K.~D. 2011, \apj, 737, 29

\bibitem[{{Papadakis} {et~al.}(2000){Papadakis}, {Hatzidimitriou}, {Croke}, \&
  {Papamastorakis}}]{Papadakis00}
{Papadakis}, I., {Hatzidimitriou}, D., {Croke}, B.~F.~W., \& {Papamastorakis},
  I. 2000, \aj, 119, 851

\bibitem[{{Peterson}(1978)}]{Peterson78}
{Peterson}, R. 1978, \apj, 222, 181

\bibitem[{{Piatek} {et~al.}(2008){Piatek}, {Pryor}, \& {Olszewski}}]{Piatek08}
{Piatek}, S., {Pryor}, C., \& {Olszewski}, E.~W. 2008, \aj, 135, 1024

\bibitem[{{Piskunov} {et~al.}(1995){Piskunov}, {Kupka}, {Ryabchikova}, {Weiss},
  \& {Jeffery}}]{Piskunov95}
{Piskunov}, N.~E., {Kupka}, F., {Ryabchikova}, T.~A., {Weiss}, W.~W., \&
  {Jeffery}, C.~S. 1995, \aaps, 112, 525

\bibitem[{{Pomp{\'e}ia} {et~al.}(2008){Pomp{\'e}ia}, {Hill}, {Spite}, {Cole},
  {Primas}, {Romaniello}, {Pasquini}, {Cioni}, \& {Smecker Hane}}]{Pompeia08}
{Pomp{\'e}ia}, L., {Hill}, V., {Spite}, M., {Cole}, A., {Primas}, F.,
  {Romaniello}, M., {Pasquini}, L., {Cioni}, M.-R., \& {Smecker Hane}, T. 2008,
  \aap, 480, 379

\bibitem[{{Preston} {et~al.}(2006){Preston}, {Sneden}, {Thompson}, {Shectman},
  \& {Burley}}]{Preston06}
{Preston}, G.~W., {Sneden}, C., {Thompson}, I.~B., {Shectman}, S.~A., \&
  {Burley}, G.~S. 2006, \aj, 132, 85

\bibitem[{{Primas} {et~al.}(2000){Primas}, {Asplund}, {Nissen}, \&
  {Hill}}]{Primas00}
{Primas}, F., {Asplund}, M., {Nissen}, P.~E., \& {Hill}, V. 2000, \aap, 364,
  L42

\bibitem[{{Robertson} {et~al.}(2005){Robertson}, {Bullock}, {Font}, {Johnston},
  \& {Hernquist}}]{Robertson05}
{Robertson}, B., {Bullock}, J.~S., {Font}, A.~S., {Johnston}, K.~V., \&
  {Hernquist}, L. 2005, \apj, 632, 872

\bibitem[{{Sadakane} {et~al.}(2004){Sadakane}, {Arimoto}, {Ikuta}, {Aoki},
  {Jablonka}, \& {Tajitsu}}]{Sadakane04}
{Sadakane}, K., {Arimoto}, N., {Ikuta}, C., {Aoki}, W., {Jablonka}, P., \&
  {Tajitsu}, A. 2004, \pasj, 56, 1041

\bibitem[{{Sandstrom} {et~al.}(2001){Sandstrom}, {Pilachowski}, \&
  {Saha}}]{Sandstrom01}
{Sandstrom}, K., {Pilachowski}, C.~A., \& {Saha}, A. 2001, \aj, 122, 3212

\bibitem[{{Sbordone} {et~al.}(2007){Sbordone}, {Bonifacio}, {Buonanno},
  {Marconi}, {Monaco}, \& {Zaggia}}]{Sbordone07}
{Sbordone}, L., {Bonifacio}, P., {Buonanno}, R., {Marconi}, G., {Monaco}, L.,
  \& {Zaggia}, S. 2007, \aap, 465, 815

\bibitem[{{Shetrone} {et~al.}(2009){Shetrone}, {Siegel}, {Cook}, \&
  {Bosler}}]{Shetrone09}
{Shetrone}, M.~D., {Siegel}, M.~H., {Cook}, D.~O., \& {Bosler}, T. 2009, \aj,
  137, 62

\bibitem[{{Short} \& {Hauschildt}(2006)}]{Short06}
{Short}, C.~I. \& {Hauschildt}, P.~H. 2006, \apj, 641, 494

\bibitem[{{Smith}(2004)}]{Smith04}
{Smith}, H.~A. 2004, {RR Lyrae Stars}

\bibitem[{{Smolec}(2005)}]{Smolec05}
{Smolec}, R. 2005, Acta Astronomica, 55, 59

\bibitem[{{Sneden}(1973)}]{Sneden73}
{Sneden}, C.~A. 1973, PhD thesis, The University of Texas at Austin

\bibitem[{{Soszy{\'n}ski} {et~al.}(2003){Soszy{\'n}ski}, {Udalski},
  {Szymanski}, {Kubiak}, {Pietrzynski}, {Wozniak}, {Zebrun}, {Szewczyk}, \&
  {Wyrzykowski}}]{Soszynski03}
{Soszy{\'n}ski}, I., {Udalski}, A., {Szymanski}, M., {Kubiak}, M.,
  {Pietrzynski}, G., {Wozniak}, P., {Zebrun}, K., {Szewczyk}, O., \&
  {Wyrzykowski}, L. 2003, Acta Astronomica, 53, 93

\bibitem[{{Soszy{\'n}ski} {et~al.}(2010){Soszy{\'n}ski}, {Udalski},
  {Szyma{\'n}ski}, {Kubiak}, {Pietrzy{\'n}ski}, {Wyrzykowski}, {Ulaczyk}, \&
  {Poleski}}]{Soszynski10b}
{Soszy{\'n}ski}, I., {Udalski}, A., {Szyma{\'n}ski}, M.~K., {Kubiak}, J.,
  {Pietrzy{\'n}ski}, G., {Wyrzykowski}, {\L}., {Ulaczyk}, K., \& {Poleski}, R.
  2010, Acta Astronomica, 60, 165

\bibitem[{{Soszy{\'n}ski} {et~al.}(2009){Soszy{\'n}ski}, {Udalski},
  {Szyma{\'n}ski}, {Kubiak}, {Pietrzy{\'n}ski}, {Wyrzykowski}, {Szewczyk},
  {Ulaczyk}, \& {Poleski}}]{Soszynski09}
{Soszy{\'n}ski}, I., {Udalski}, A., {Szyma{\'n}ski}, M.~K., {Kubiak}, M.,
  {Pietrzy{\'n}ski}, G., {Wyrzykowski}, {\L}., {Szewczyk}, O., {Ulaczyk}, K.,
  \& {Poleski}, R. 2009, Acta Astronomica, 59, 1

\bibitem[{{Stellingwerf}(1978)}]{Stellingwerf78}
{Stellingwerf}, R.~F. 1978, \apj, 224, 953

\bibitem[{{Tafelmeyer} {et~al.}(2010){Tafelmeyer}, {Jablonka}, {Hill},
  {Shetrone}, {Tolstoy}, {Irwin}, {Battaglia}, {Helmi}, {Starkenburg}, {Venn},
  {Abel}, {Francois}, {Kaufer}, {North}, {Primas}, \&
  {Szeifert}}]{Tafelmeyer10}
{Tafelmeyer}, M., {Jablonka}, P., {Hill}, V., {Shetrone}, M., {Tolstoy}, E.,
  {Irwin}, M.~J., {Battaglia}, G., {Helmi}, A., {Starkenburg}, E., {Venn},
  K.~A., {Abel}, T., {Francois}, P., {Kaufer}, A., {North}, P., {Primas}, F.,
  \& {Szeifert}, T. 2010, \aap, 524, A58

\bibitem[{{Thielemann} {et~al.}(1996){Thielemann}, {Nomoto}, \&
  {Hashimoto}}]{Thielemann96}
{Thielemann}, F.-K., {Nomoto}, K., \& {Hashimoto}, M.-A. 1996, \apj, 460, 408

\bibitem[{{Tumlinson}(2010)}]{Tumlinson10}
{Tumlinson}, J. 2010, \apj, 708, 1398

\bibitem[{{Udalski} {et~al.}(1997){Udalski}, {Kubiak}, \&
  {Szymanski}}]{Udalski97}
{Udalski}, A., {Kubiak}, M., \& {Szymanski}, M. 1997, Acta Astronomica, 47, 319

\bibitem[{{Udalski} {et~al.}(2008{\natexlab{a}}){Udalski}, {Soszy{\'n}ski},
  {Szymanski}, {Kubiak}, {Pietrzynski}, {Wyrzykowski}, {Szewczyk}, {Ulaczyk},
  \& {Poleski}}]{Udalski08a}
{Udalski}, A., {Soszy{\'n}ski}, I., {Szymanski}, M.~K., {Kubiak}, M.,
  {Pietrzynski}, G., {Wyrzykowski}, L., {Szewczyk}, O., {Ulaczyk}, K., \&
  {Poleski}, R. 2008{\natexlab{a}}, Acta Astronomica, 58, 89

\bibitem[{{Udalski} {et~al.}(2008{\natexlab{b}}){Udalski}, {Soszy{\'n}ski},
  {Szyma{\'n}ski}, {Kubiak}, {Pietrzy{\'n}ski}, {Wyrzykowski}, {Szewczyk},
  {Ulaczyk}, \& {Poleski}}]{Udalski08b}
{Udalski}, A., {Soszy{\'n}ski}, I., {Szyma{\'n}ski}, M.~K., {Kubiak}, M.,
  {Pietrzy{\'n}ski}, G., {Wyrzykowski}, {\L}., {Szewczyk}, O., {Ulaczyk}, K.,
  \& {Poleski}, R. 2008{\natexlab{b}}, Acta Astronomica, 58, 329

\bibitem[{{Valdes} {et~al.}(2004){Valdes}, {Gupta}, {Rose}, {Singh}, \&
  {Bell}}]{Valdes04}
{Valdes}, F., {Gupta}, R., {Rose}, J.~A., {Singh}, H.~P., \& {Bell}, D.~J.
  2004, \apjs, 152, 251

\bibitem[{{van der Marel} {et~al.}(2009){van der Marel}, {Kallivayalil}, \&
  {Besla}}]{Marel09}
{van der Marel}, R.~P., {Kallivayalil}, N., \& {Besla}, G. 2009, in IAU
  Symposium 256, ed. J.~T. {van Loon} \& J.~M. {Oliveira} (Cambridge University
  Press, Cambridge)

\bibitem[{{Venn} {et~al.}(2004){Venn}, {Irwin}, {Shetrone}, {Tout}, {Hill}, \&
  {Tolstoy}}]{Venn04}
{Venn}, K.~A., {Irwin}, M., {Shetrone}, M.~D., {Tout}, C.~A., {Hill}, V., \&
  {Tolstoy}, E. 2004, \aj, 128, 1177

\bibitem[{{Venn} {et~al.}(2012){Venn}, {Shetrone}, {Irwin}, {Hill}, {Jablonka},
  {Tolstoy}, {Lemasle}, {Divell}, {Starkenburg}, {Letarte}, {Baldner},
  {Battaglia}, {Helmi}, {Kaufer}, \& {Primas}}]{Venn12}
{Venn}, K.~A., {Shetrone}, M.~D., {Irwin}, M.~J., {Hill}, V., {Jablonka}, P.,
  {Tolstoy}, E., {Lemasle}, B., {Divell}, M., {Starkenburg}, E., {Letarte}, B.,
  {Baldner}, C., {Battaglia}, G., {Helmi}, A., {Kaufer}, A., \& {Primas}, F.
  2012, \apj, 751, 102

\bibitem[{{Vivas} {et~al.}(2005){Vivas}, {Zinn}, \& {Gallart}}]{Vivas05}
{Vivas}, A.~K., {Zinn}, R., \& {Gallart}, C. 2005, \aj, 129, 189

\bibitem[{{Wallerstein} {et~al.}(1963){Wallerstein}, {Greenstein}, {Parker},
  {Helfer}, \& {Aller}}]{Wallerstein63}
{Wallerstein}, G., {Greenstein}, J.~L., {Parker}, R., {Helfer}, H.~L., \&
  {Aller}, L.~H. 1963, \apj, 137, 280

\bibitem[{{Woosley} \& {Weaver}(1995)}]{Woosley95}
{Woosley}, S.~E. \& {Weaver}, T.~A. 1995, \apjs, 101, 181

\bibitem[{{Wyse} {et~al.}(2002){Wyse}, {Gilmore}, {Houdashelt}, {Feltzing},
  {Hebb}, {Gallagher}, \& {Smecker-Hane}}]{Wyse02}
{Wyse}, R.~F.~G., {Gilmore}, G., {Houdashelt}, M.~L., {Feltzing}, S., {Hebb},
  L., {Gallagher}, III, J.~S., \& {Smecker-Hane}, T.~A. 2002, New Astronomy, 7,
  395

\bibitem[{{Yoon} \& {Lee}(2002)}]{Yoon02}
{Yoon}, S.-J. \& {Lee}, Y.-W. 2002, Science, 297, 578

\bibitem[{{Zinn} \& {West}(1984)}]{Zinn84}
{Zinn}, R. \& {West}, M.~J. 1984, \apjs, 55, 45

\end{thebibliography}
\bibliographystyle{apj}

\newpage
\appendix

\begin{figure}[h!]
\includegraphics[width=1\textwidth]{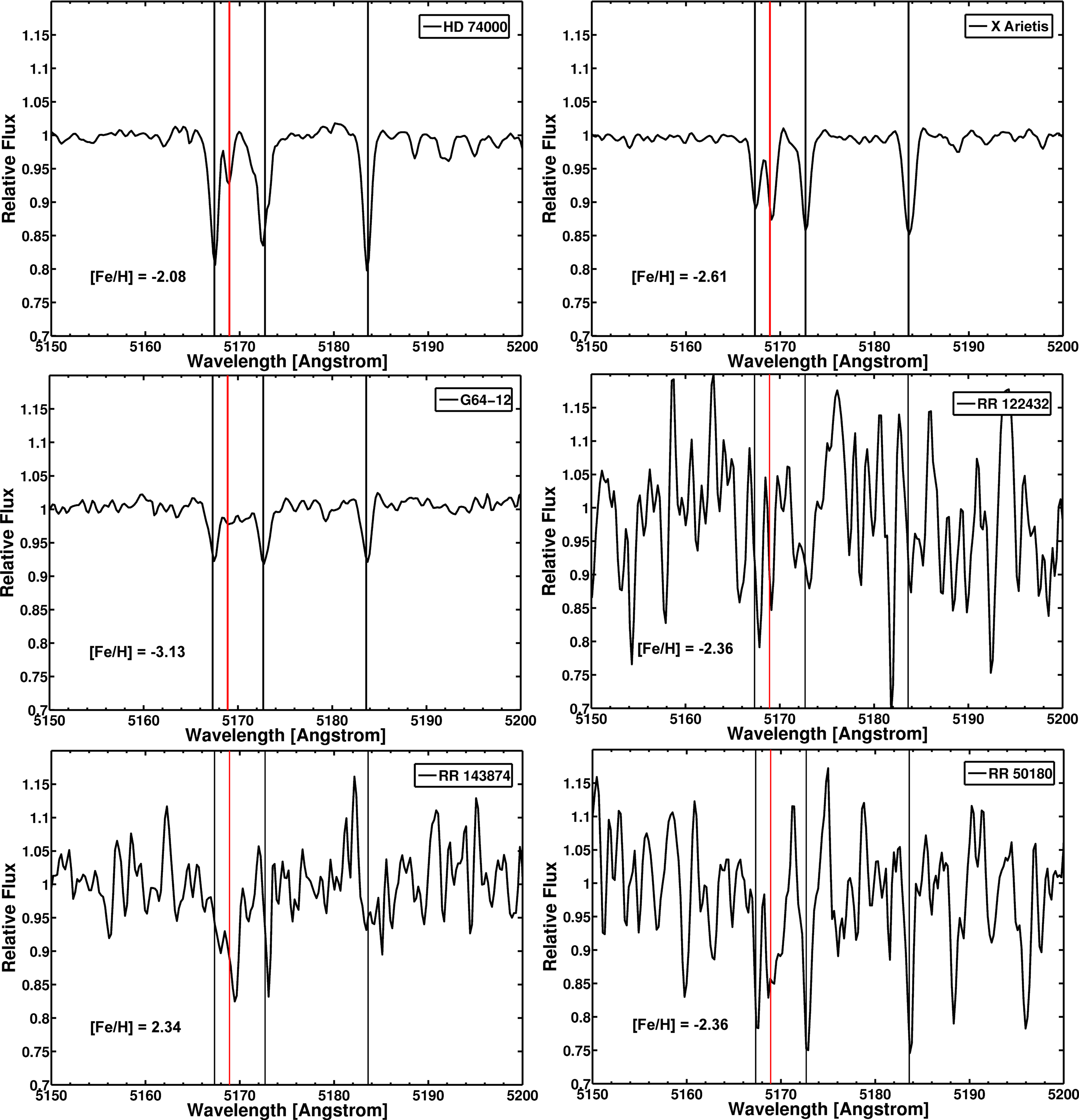} 
\caption{Spectra of the Mg lines between 5150\AA \ and 5200\AA. The black lines show the Mg lines, while the red lines show an Fe\,I line. The standard stars HD74000, G64-12 and X\,Ari and our three SMC targets are shown.} 
\label{spectra_Mg_standards_SMC}
\end{figure}

\begin{figure}
\includegraphics[width=1\textwidth]{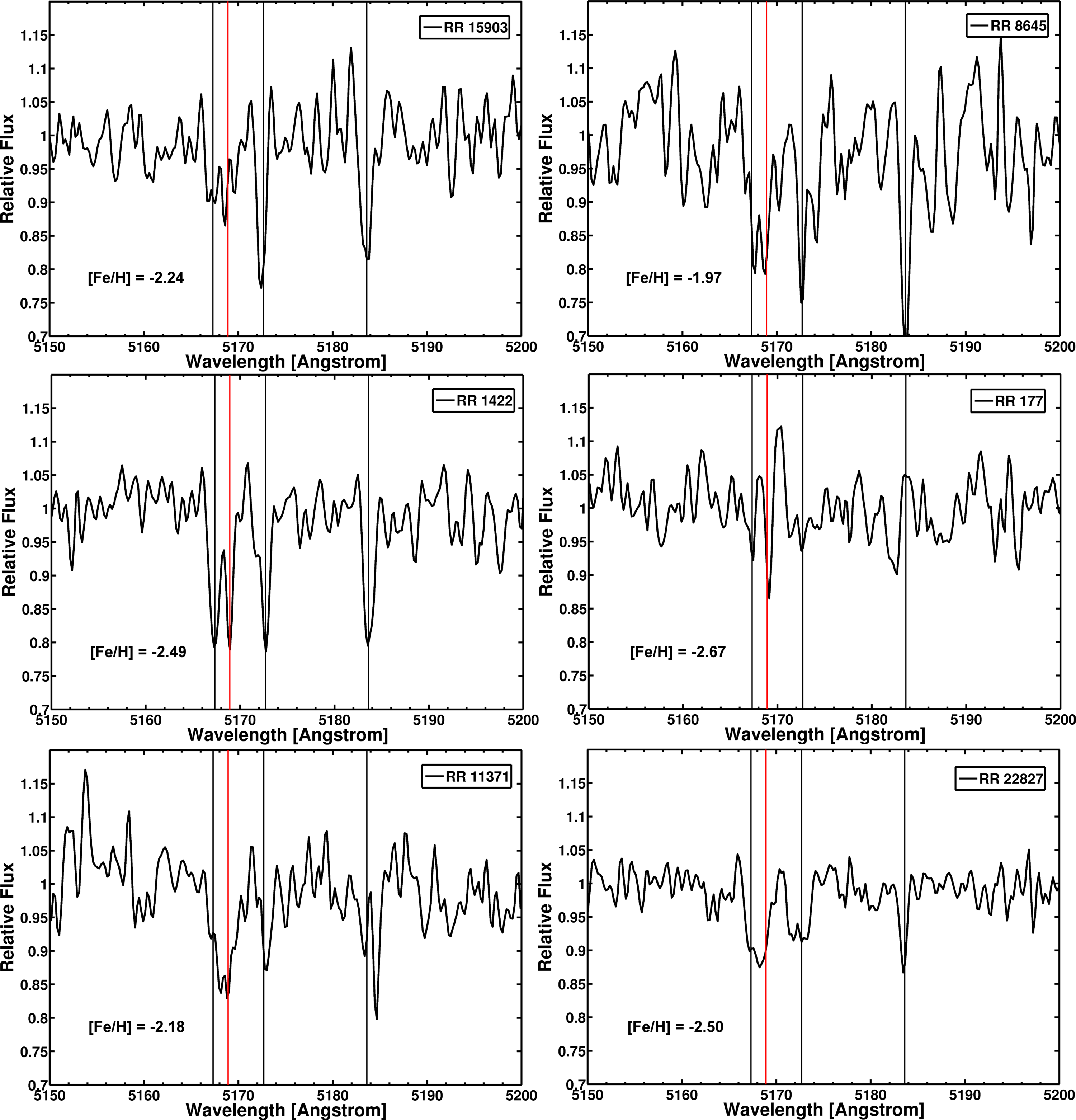} 
\caption{Spectra of the Mg lines between 5150\AA \ and 5200\AA. The black lines show the Mg lines, while the red lines show an Fe\,I line. All targets from the LMC are shown here.} 
\label{spectra_Mg_LMC}
\end{figure}

\end{document}